\documentclass[%
prx,twocolumn,showpacs,superscriptaddress,nobibnotes,
 amsmath,amssymb,nofootinbib,
 aps,
]{revtex4-2}

\usepackage{graphicx}
\usepackage{xcolor}
\usepackage[hidelinks]{hyperref}
\usepackage{dcolumn}
\usepackage{bm}

\begin{document}


\title{Characterizing nonlinear dynamics by contrastive cartography}

\author{Nicolas Romeo}
\email{nromeo@uchicago.edu}
\affiliation{%
 Center for Living Systems, University of Chicago, Chicago, Illinois 60637
}%
\affiliation{%
 Department of Physics, University of Chicago, Chicago, Illinois 60637
}%
\author{Chris Chi}%
\affiliation{%
 Center for Living Systems, University of Chicago, Chicago, Illinois 60637
}%
\affiliation{%
 Department of Chemistry, University of Chicago, Chicago, Illinois 60637
}%
 

\author{Aaron R. Dinner}
\email{dinner@uchicago.edu}
\affiliation{%
 Center for Living Systems, University of Chicago, Chicago, Illinois 60637
}%
\affiliation{%
 Department of Chemistry, University of Chicago, Chicago, Illinois 60637
}%
\author{Elizabeth R. Jerison}
\email{ejerison@uchicago.edu}
\affiliation{%
 Center for Living Systems, University of Chicago, Chicago, Illinois 60637
}%
\affiliation{%
 Department of Physics, University of Chicago, Chicago, Illinois 60637
}%

\date{\today}

\begin{abstract}

The qualitative study of dynamical systems using bifurcation theory is key to understanding systems from biological clocks and neurons to physical phase transitions. Data generated from such systems can feature complex transients, an unknown number of attractors, and stochasticity.  Characterization of these often-complicated behaviors remains challenging. Making an analogy to bifurcation analysis, which specifies that useful dynamical features are often invariant to coordinate transforms, we leverage contrastive learning to devise a generic tool to discover dynamical classes from stochastic trajectory data. By providing a model-free trajectory analysis tool, this method automatically recovers the dynamical phase diagram of known models and provides a ``map'' of dynamical behaviors for a large ensemble of dynamical systems. The method thus provides a way to characterize and compare dynamical trajectories without governing equations or prior knowledge of target behavior. We additionally show that the same strategy can be used to characterize the stochastic motion of bacteria, establishing that this approach can be used as a standalone analysis tool or as a component of a broader data-driven analysis framework for dynamical data.

\end{abstract}

\maketitle

Complex dynamics in many biological, physical, and engineered systems are difficult to analyze: even relatively simple nonlinear dynamical systems can display markedly different behaviors as their parameters are changed, from oscillations and spikes to chaos \cite{strogatz_nonlinear_2024,izhikevich_dynamical_2006, kuznetsov_elements_2023,meiss_differential_2007}. With the emergence of powerful data-driven approaches, symbolic or generative models fit to data are capable of reproducing experimentally observed time series \cite{gilpin_generative_2024,brunton_discovering_2016,supekar_learning_2023,schmidt_distilling_2009,lu_extracting_2020, frishman_learning_2020,kadeethum_framework_2021}. However, it remains challenging to identify and interpret the qualitative behaviors that such complex and often nonlinear models generate \cite{falk_curiosity-driven_2024}. For example, suppose you observed a biochemical system that oscillated in some regimes, and was multistable in others. How would you identify those behaviors?

The classical tool to study behavior regimes of nonlinear dynamical systems is bifurcation theory.
Concerned with the study of qualitative changes in the flow of dynamical systems, it is the basis of theories across physical and biological disciplines, from the dynamical picture of phase transitions~\cite{fruchart_non-reciprocal_2021,tauber_critical_2014} and the analysis of biological regulatory networks~\cite{elowitz_synthetic_2000,alon_introduction_2020}, to the identification of ecological tipping points~\cite{dai_generic_2012, dakos_ecosystem_2019} and reduced order modeling in engineering~\cite{brunton_data-driven_2022}.

Through the characterization of attractors in the dynamical flow, bifurcation theory provides a link between model parameters and behaviors. Defining observables (order parameters) characteristic of these attractors allows for quantification of changes in target behavior. 
This quantification in principle allows for determining which behavior regimes are present in experimental and observational data.
However, in practice, differences between behaviors can be subtle, necessitating the development and use of specific statistical tests for each behavior of interest.
For example, extensive work has been devoted to determining when oscillations are present in time series data \cite{welch_use_1967,dickey_distribution_1979,hutchison_improved_2015}. Additionally, most studies of bifurcations are limited to determining the invariant sets and linear analysis of flows around them. 

However, global bifurcations involve the interaction of multiple flow structures and cannot be captured through linear analysis around invariant sets, making the definition of observables much more difficult \cite{hoppensteadt_weakly_1997,izhikevich_neural_2000}. Defining summary statistics that capture complex transient behaviors is also challenging \cite{hastings_transient_2018, koch_ghost_2024}.
The presence of noise further complicates analyses based on bifurcation theory, as the definition of invariant sets and flows requires the introduction of probabilistic criteria \cite{kwon_structure_2005, billings_identifying_2008,dickey_distribution_1979,chung_fully_2017}.

These challenges make developing statistics for complex behaviors a difficult task.
These hurdles arise often in analyses of living systems, where complex, noisy dynamics are common, and there is usually no first-principles model of the dynamics.
To define behavior without the need for prior knowledge, recent work has developed generic ways to characterize nonlinear dynamical systems by studying their approximations via linear models \cite{costa_adaptive_2019,romeo_learning_2021,cohen_schrodinger_2023} or Markov chains \cite{wiltschko_mapping_2015,sridhar_uncovering_2024}.  While powerful and adaptable, these approaches have the drawback that they can obscure simple nonlinear effects in favor of complex linear models. For example, no linear model can be bistable.
 
Machine-learning methods for reducing the dimensionality of dynamical data to a few informative variables have proven effective to both simplify the dynamical space \cite{schmitt_information_2024,conti_veni_2024, zeng_autoencoders_2024}, and to characterize complex time-series \cite{falk_curiosity-driven_2024,achar_universal_2022,berman_mapping_2014}. These methods take data generated by a particular process---either a model or an experiment---and train a neural network to produce a meaningful low-dimensional embedding. While powerful, the characterizations obtained from these methods do not transparently connect to bifurcation theory, making the study of transitions difficult.

Here, our goal is to characterize the dynamical behavior of a set of stochastic trajectories without prior knowledge, in a manner that recovers bifurcation theory when it is applicable. One earlier study also pursued this goal for deterministic trajectories \cite{yair_reconstruction_2017}. Their approach, based on diffusion maps, yielded promising results on a small number of model systems, but necessitated complex hand-construction of pre-processing and dimensional reduction steps, and used methods that do not allow for out-of-sample comparisons. Given advances in machine learning, we were motivated to ask whether a different approach could enable application to noisy trajectories across a broad range of nonlinear systems.

Our strategy is to use contrastive learning, which is commonly used in modern computer vision and natural language processing \cite{oord_representation_2019,chen_simple_2020,chuang_dfiffcse_2022,loh_surrogate_2022}, to learn a representation of trajectories characterizing the observed dynamics. By designing data augmentations based on some of the topological invariances of these systems, we recover a latent space that is consistent with bifurcation theory and can be used to classify and separate dynamical behaviors. We further analyze experimental data for a biological process for which low-dimensional dynamical bifurcations are less relevant because of intrinsic stochasticity and complex dynamics that are internal to cells, and we show that a similar self-supervised strategy to incorporate geometric invariance can still provide more informative latent spaces than generic dimensional reduction methods.

\section*{Results}

We present the construction of the latent space and evaluate its ability to discriminate between stereotypical dynamical behaviors. We then show that it reflects the topological structure of the dynamical flows that generated the input data and present additional features of the latent space that can make the method useful for general dynamical data exploration. We finally show, using an application to run-and-tumble bacterial tracks, that accounting for invariances using a similar contrastive strategy can be valuable to characterize strongly stochastic dynamics.

\subsection*{Contrastive learning framework}


\begin{figure*}
    \centering
    \includegraphics[width=\linewidth]{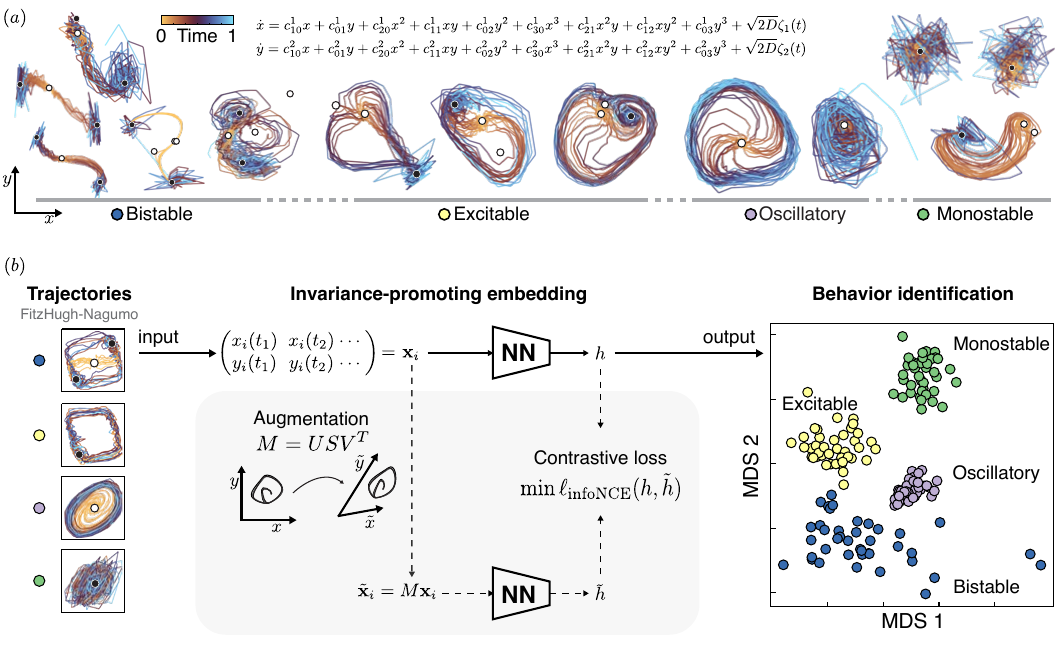}
    \caption{\textbf{Contrastive learning to identify dynamical behavior.} \emph{(a)} A sample of trajectories generated from an ensemble of 2D dynamical systems with polynomial nonlinearites. The systems display a variety of dynamical behaviors. Color indicates time; white symbols indicate unstable fixed points; black symbols indicate stable fixed points.  \emph{(b)} Proposed pipeline for identifying dynamical behaviors, here applied to trajectories generated by the FitzHugh-Nagumo model with various parameter choices. During training (on the dataset of \emph{(a)}), linear transformations are applied to input trajectories to generate an augmented dataset which allows the optimization of a contrastive loss (dashed lines). Collections of trajectories $\{\mathbf{x}(t_i)\}$ are embedded into a smaller-dimensional space $h$, in which different dynamical behaviors emerge as clusters as seen through projection into a 2D MDS basis.}
    \label{fig:constrastive}
\end{figure*}

Dynamical systems can exhibit various behaviors. 
Consider a two-dimensional (2D) stochastic dynamical system $\dot{\mathbf{x}}=\mathbf{f}(\mathbf{x}) + \boldsymbol{\zeta}(t)$ that governs the evolution of a variable $\mathbf{x}= (x,y)$ under a deterministic flow $\mathbf{f}$ and standard Brownian noise $\boldsymbol{\zeta}(t)$ with variance $D$.
A generic 2D smooth $\mathbf{f}$ can be Taylor-expanded to order $N$ as 
\begin{equation}
    \mathbf{f}(\mathbf{x}) = \sum_{k=1,2}\sum_{0 \leq i+j \leq N} c^k_{ij} x^i y^j \mathbf{e}_k + O(x^{N+1}, y^{N+1}) \label{eq:ode_TaylorExpansion}
\end{equation}
with $\mathbf{e}_1 = (1,0)^T$ and $\mathbf{e}_2 = (0,1)^T$ the 2D Cartesian basis vectors. For $N=3$, the resulting cubic-order differential equations cover many commonly encountered physical and biological models described by cubic normal forms and canonical models \cite{strogatz_nonlinear_2024, izhikevich_dynamical_2006}. 
As we simulate an ensemble of such systems with normally sampled coefficients $c^k_{ij}$ up to $N=3$, under the restriction that $\mathbf{f}\cdot \mathbf{x} \leq 0$ as $\mathbf{x}\rightarrow \infty$ to guarantee boundedness of the flow (SI Sec.~I), we observe a diversity of commonly encountered phenomena such as (multi-)stability, oscillations, or excitable spiking transients, with smooth transitions between them (Fig.~\ref{fig:constrastive}(a)). While the different 2D dynamical behaviors are clear to the eye away from transition regimes, they are hard to characterize directly from trajectory data without prior knowledge of desired characteristics. Similar arguments apply to $d$-dimensional dynamical systems for which cubic polynomial systems can exhibit even more complex behaviors, including chaos in $d\geq3$ \cite{strogatz_nonlinear_2024,meiss_differential_2007}.

A core insight from bifurcation theory is that the topological features of the flow $\mathbf{f}$ provide a way to classify dynamical systems, most commonly by establishing the presence, number, and stability of fixed points and attractors \cite{kuznetsov_elements_2023,meiss_differential_2007,izhikevich_dynamical_2006}. Such topological objects are invariant under any continuous change of coordinate system (or homeomorphism), and traditional normal form analysis seeks to find particular coordinates in which dynamics are reduced to well-characterized normal forms with particularly simple flows of the form given by Eq.~\eqref{eq:ode_TaylorExpansion}. This perspective suggests that a neural network trained to recognize features invariant to coordinate changes should be able to detect what we mean by dynamical behavior. 

To implement this idea, we use a self-supervised encoder neural network which takes as input sets of regularly-sampled trajectories $\{\mathbf{x}(t)\}$ and outputs a vector $h[\{\mathbf{x}(t)\}] \in \mathbb{R}^n $ characterizing the dynamical system. The central object of this article is the learned embedding map
\begin{equation}
    h: \{\mathbf{x}(t)\} \mapsto h[\{\mathbf{x}(t)\}],
\end{equation} 
which we dub the \emph{cartographer}. Our goal is to construct and train the cartographer to be interpretable in light of bifurcation theory by training it to recognize coordinate-invariant features of the input data.
The construction of a neural network whose input is invariant to specific transformations, known as \emph{augmentations}, is the central idea behind contrastive learning~\cite{chen_simple_2020}: The name contrastive learning reflects that the trained cartographer $h$ must separate samples that are related by an augmentation transformation from ones that are not (SI Sec.~II). Such augmented samples should be clustered together in the embedding space, thus imposing a soft invariance constraint. However, arbitrary continuous coordinate transformations by construction erase all geometric data that could provide a basis to distinguish between otherwise topologically equivalent systems. We thus restrict ourselves to invertible linear coordinate changes along with permutations of the input trajectories (SI Sec.~II.B).

Having identified a class of suitable augmentation transformations, we optimize the InfoNCE loss function, a common choice for contrastive learning with strong ties to information theory \cite{oord_representation_2019,gokmen_symmetries_2021, murphy_machine-learning_2024, schmitt_information_2024,jiang_training_2023}.  Details of the neural network architecture and loss function are provided in SI Sec.~II.A.
Our last task to construct the map $h$ is to define an appropriate training dataset that spans a diverse set of dynamics; for this, we integrate $2\cdot10^4$ systems sampled using the same protocol as for Fig.~\ref{fig:constrastive}(a) (SI Sec.~I). With these training data, we find that our neural network can discriminate between different parameter regimes of stochastic FitzHugh-Nagumo systems with noise amplitude $D=0.1$ (Fig.~\ref{fig:constrastive}(b), SI Sec.~III.A). In the supplementary materials, we investigate the effect of varying initial conditions (SI Sec.~VI) and compare our method to an autocorrelation-based approach, which, while simpler and training-free, is less accurate and does not provide an invariant characterization (SI Sec.~VII, Fig.~S10). Compared to characterizing dynamics by finding sequences of linear propagators \cite{costa_adaptive_2019}, our approach is technically simpler and does not require finding a link between piecewise linear models and nonlinear behavior.


\begin{figure*}
    \centering
    \includegraphics[width=.85\linewidth]{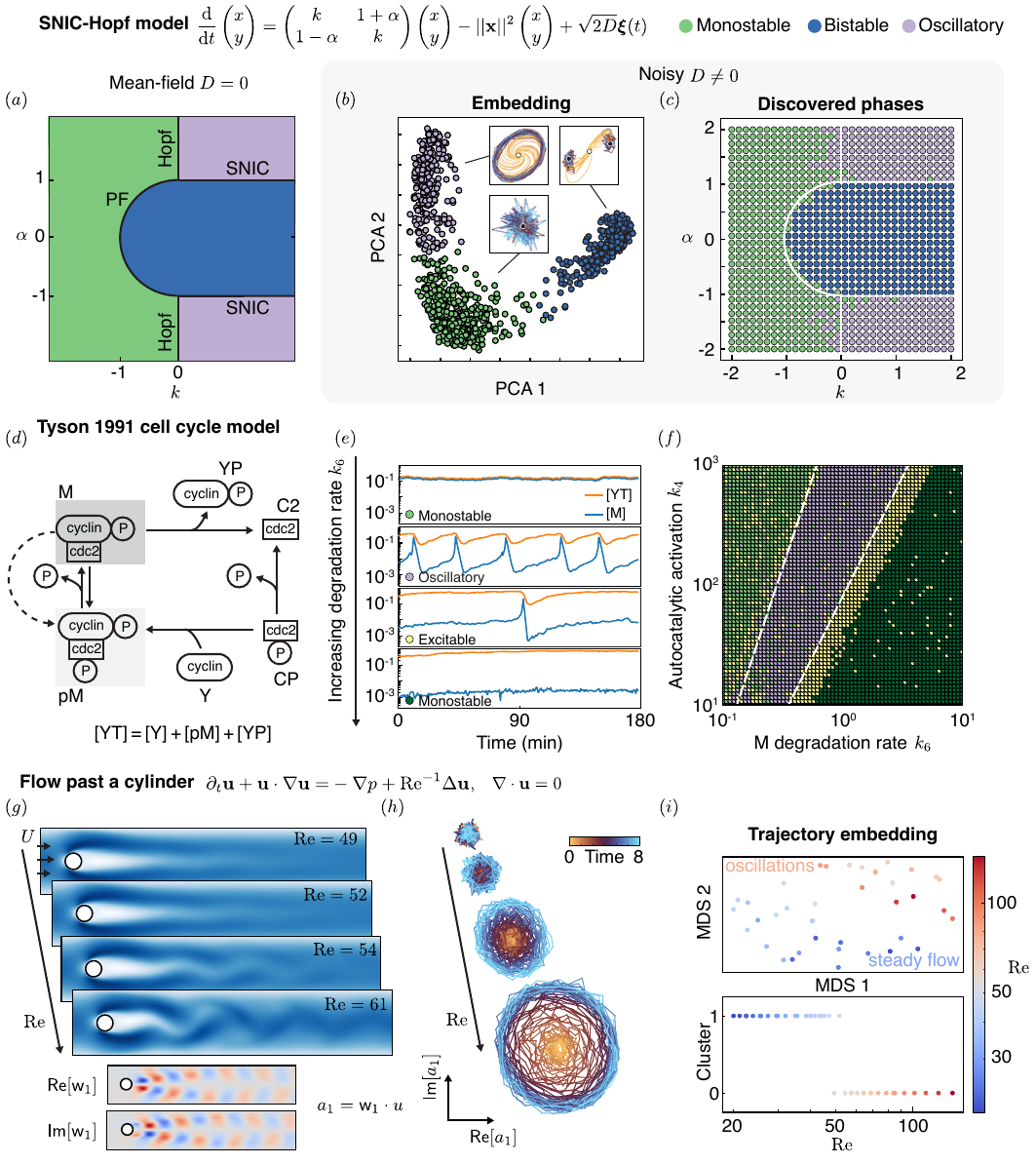}
    \caption{\textbf{Dynamical phase identification in example 2D systems.}  \emph{(a)} The SNIC-Hopf model system displays multiple possible behaviors as the parameters $k$ and $\alpha$ are changed. Analytical results show that the different phases are separated by Pitchfork (PF), Hopf and Saddle-Node on an Invariant Circle (SNIC) bifurcations. \emph{(b)} A 3-means clustering in the contrastive embedding space recovers the main distinct dynamics. Inserts show typical trajectories in the $(x,y)$ plane. \emph{(c)} Inferred phase diagram using the identified phase labels from \emph{(b)}. Mean-field results are overlaid in white. \emph{(d)} A six-dimensional cell cycle model \cite{tyson_modeling_1991}. \emph{(e)} Varying the degradation rate of the cdc2-cylin-P complex (denoted M) leads to distinct dynamical behaviors in the presence of noise. Concentrations are expressed in units of total cdc2 concentration. \emph{(f)} Agglomerative clustering in latent space recovers the boundaries of the distinct behaviors in accordance with previous analytical results (white lines, see SI and \cite{tyson_modeling_1991}). \emph{(g)} As intake velocity $U$ is increased, the flow past a channel displays a von K\'arm\'an vortex street and becomes oscillatory in time  \emph{(h)} A noisy Galerkin projection of the flow $\mathbf{u}$ onto $\mathsf{w}_1$, the first dynamic mode at Reynolds number $\mathrm{Re}= UL/\nu = 100$, shows the emergence of a limit cycle. \emph{(i)} Contrastive embedding of reduced time series shows the emergence of two clusters separated at $Re\approx 50$, providing a characterization of the dynamics.}
    \label{fig:phasediagrams}
\end{figure*}

\subsection*{Cartographer classifies behaviors}

What kind of understanding does the cartographer provide about dynamical systems?
To answer this question, we now apply the cartographer trained on the data generated by cubic polynomial systems [i.e., the data described in the previous section and illustrated in Fig.\ \ref{fig:constrastive}(a)] to a series of examples.  We emphasize that we do not further train the cartographer for any example. 

As a first example, we consider the problem of identifying distinct dynamical behaviors without prior assumptions, on a system for which an analytical phase diagram is known. 
We thus consider a minimal model of non-reciprocal dynamics arising in active matter and ecological systems, dubbed the \emph{SNIC-Hopf model}, given by \cite{martin_transition_2024}
\begin{align}
    \frac{\mathrm{d}}{\mathrm{d}t}\begin{pmatrix}
        x \\ y
    \end{pmatrix} = \begin{pmatrix}
        k & 1+ \alpha \\ 1-\alpha & k
    \end{pmatrix} \begin{pmatrix}
        x \\ y
    \end{pmatrix}  - (x^2+y^2) \begin{pmatrix}
        x \\ y
    \end{pmatrix}. \label{eq:SNIChopf}
\end{align}
 In the absence of noise $D=0$, this system can display three different behaviors separated by well-characterized bifurcations (Fig.~\ref{fig:phasediagrams}(a), see SI Sec.~III.B). We numerically integrate the corresponding stochastic differential equations (SDEs) for varying $k$ and $\alpha$ and fixed non-zero noise $D=5\cdot10^{-4}$, and embed the trajectories using our contrastive map [Fig.~\ref{fig:phasediagrams}(b)]. The embedding visually reveals three distinct regions. Using $K$-means clustering on the embedded data $\{h\}$ with $K=3$, we recover the dynamical phases and their expected boundaries, albeit slightly shifted [Fig.~\ref{fig:phasediagrams}(c)]. These shifts are expected: oscillations are observed even in the monostable regime in the vicinity of Hopf bifurcations in noisy systems, a phenomenon known as noise-induced oscillations \cite{vilar_mechanisms_2002}, while the presence of noise tends to delay the visibility of bistability, as seen by the misidentification of bistable systems as monostable near the transition \cite{gaspard_spectral_1995}. An additional example featuring more complex dynamical flow features is presented in SI Sec.~III.C, Fig.~S3: we find there that global bifurcations, which are discontinuous in flow space, are often easier to accurately locate than local bifurcations.

As a more complex example, we consider a six-dimensional dynamical model of the cell cycle presented in \cite{tyson_modeling_1991}; we focus on the dynamics of the cyclin and cdc2 proteins as they get phosphorylated and bind each other (Fig.~\ref{fig:phasediagrams}(d); see SI Sec.~III.D for model details). Under variation of the degradation rate $k_6$ of the phosphorylated complex $\mathrm{M}$, the model can display distinct steady states separated by oscillatory and excitable regimes [Fig.~\ref{fig:phasediagrams}(e)]. Given a two-dimensional projection of the integrated trajectories, the cartographer is able to identify these distinct regimes [Fig.~\ref{fig:phasediagrams}(f)] even though the underlying dynamics are not in the training data [Fig.\ \ref{fig:constrastive}]. We note that for applications to data for which the number of distinct behaviors is not visually clear, general clustering techniques  can be used to decide on a number of clusters to identify behaviors of interest \cite{murphy_probabilistic_2022}.

The cartographer can also be applied to data generated by a higher-dimensional process. Consider incompressible hydrodynamic flow in a channel past a circular obstacle of diameter $L$ (Fig.~\ref{fig:phasediagrams}(g); implementation details in SI Sec.~III.E). For a fluid of kinematic viscosity $\nu$, as the intake flow velocity $U$ increases, the flow undergoes a laminar-to-oscillatory transition when the Reynolds number $\mathrm{Re} = UL/\nu$ crosses the critical $\mathrm{Re}_c \approx 50$. Although this partial differential equation (PDE) problem is high-dimensional, projecting the flow on the standard basis from dynamic mode decomposition \cite{schmid_dynamic_2010} allows us to obtain a 2D time series which captures the two dynamical phases, and the cartographer is able to identify the critical Reynolds number (Fig.~\ref{fig:phasediagrams}(h-i), SI Fig.~S4). We thus see that $h$ is able to identify dynamical behaviors on data originally generated from a high-dimensional nonstationary system, suggesting potential applicability for the characterization of other PDE solutions and imaging data after suitable dimensionality reduction.


\subsection*{Cartographer recovers flow topology}

The cartographer is thus able to discriminate between common stereotypical dynamical behaviors. However, given our learning strategy, we can expect more powerful structures to emerge in the latent space encoded by $h$.  In this section, we show that the latent space trained only on trajectory data is consistent with bifurcation theory.

\begin{figure*}
    \centering
    \includegraphics[scale=1]{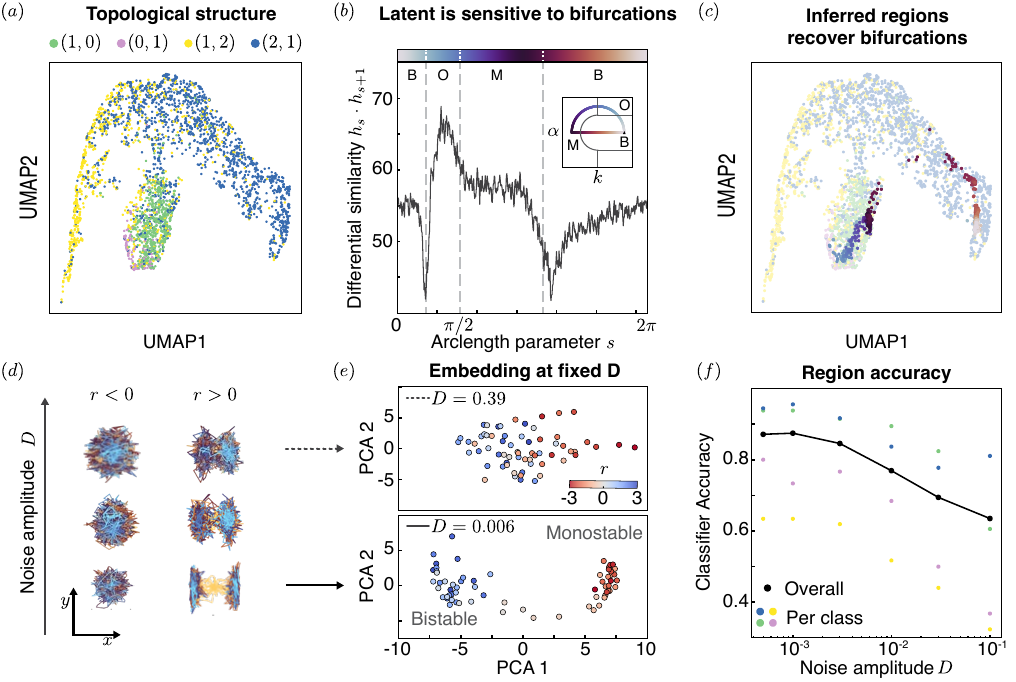}
    \caption{\textbf{The contrastive map recovers bifurcation theory at low noise.} \emph{(a)} UMAP embedding of $\sim$$2300$ samples from the random ensemble, showing clustering by number and type of fixed points $(n_\text{stable},n_\text{unstable})$.  \emph{(b)} The SNIC-Hopf system is taken along the path shown in the inset, and we examine the embedding $h(s)$ across different phases: bistable (B), oscillatory (O), and monostable (M). The embedding $h$ shows discontinuities in similarity at bifurcations. \emph{(c)} Comparing the embedding of the data in \emph{(b)} with boundaries between topological regions shows that the bifurcations match expected region boundaries.  \emph{(d)} A simple potential system exhibits a pitchfork bifurcation as the control parameter $r$ is changed, with examples at $r=-2.5$ (left) and $r=2.6$ (right). \emph{(e)} For small noise (bottom, $D=0.006$), the two distinct phases are easily identified in the latent space, while, for large noise (top, $D=0.39$), the phases are no longer separated in the latent space but the first principal component is still related to $r$.  \emph{(f)} The accuracy of a linear classifier to determine the topological structure through $h$ decreases as noise increases, in accordance with results in \emph{(d-e)}. }
    \label{fig:topo_latent}
\end{figure*}

To this end, we examine the embedding of a sample of data generated from the random dynamical ensemble (but not used to train the network). Labeling trajectories by the numerically determined fixed-point structure of the generating equations, we find that samples in latent space are clustered according to the number and stability of their fixed points (Fig.~\ref{fig:topo_latent}(a); alternative visualizations and effects of different input datasets are presented in the SI Sec.~V.D, Fig.~S8). Here, we restrict ourselves to systems with up to two fixed points of a given type. Note that our cubic systems can have up to nine fixed points; we present sampling statistics and results for more complex topologies in the SI (Sec.~V.B, Fig.~S6). Note that a complete topological characterization would also include the number and type of limit cycles, which is numerically challenging.

To explore the boundaries between dynamical regimes drawn by $h$, we follow the embedding $h(s)$ of trajectories from the SNIC-Hopf model as its parameters vary along the parametrized closed curve $\left(k(s), \alpha(s)\right)$ with $s\in[0,2\pi]$ (Fig.~S2). By measuring the variation in embedding vector between successive parameter values through the differential similarity measure $h(s_i) \cdot h(s_{i+1})$, we observe stronger variations in the vicinity of bifurcations, with transition points shifted as before from noise-induced effects [Fig.~\ref{fig:topo_latent}(b)]. Moreover, in the 2D projection shown, we indeed observe piecewise continuous paths [Fig.~\ref{fig:topo_latent}(c)].

The latent space thus allows the definition of regions associated with different dynamical phases, in the sense of fixed-point topologies (SI Sec.~V, Fig.~S7). In effect, our contrastive learning approach allows the construction of an approximate phase diagram for generic 2D dynamical systems, allowing for the identification of bifurcations by monitoring the variations in embedding.

That said, we note that there is substantial overlap between certain classes in Fig.~\ref{fig:topo_latent}(a), in apparent disagreement with the idea that flow topology determines behavior. As we now show, this is due to the stochastic nature of our systems, which breaks the clear-cut classification of deterministic dynamical systems by the nature of their asymptotic invariants. We also note that even in deterministic systems, the presence of non-normality \cite{trefethen_hydrodynamic_1993} or ``ghosts'' \cite{strogatz_nonlinear_2024,hastings_transient_2018,koch_ghost_2024,sciammas2011incoherent} can lead to distinct transient behaviors not directly captured by fixed-point topology.

The effect of noise is best understood by an example.  Consider the deterministic dynamics $\dot{x}=rx-x^3, \dot{y}=-y$. This system exhibits a single fixed point at the origin for $r<0$, while it has two stable fixed points at $(x,y)=(\pm\sqrt{r},0)$ for $r>0$. As we increase the noise, the difference between these two phases is progressively blurred [Fig.~\ref{fig:topo_latent}(d)], and this is reflected in the latent space: for small noise, as we vary $r$, the embedding presents two distinct clusters corresponding to mono- and bistable regimes, while, for large noise, there are no longer distinct clusters but the first principal component still relates to $r$ [Fig.~\ref{fig:topo_latent}(e)]. The effect of variable time integration length is discussed in SI Sec.~IV.

Coming back to the systems generated from the training ensemble, for a given trained cartographer, the accuracy of a classifier trained and tested on data integrated with noise amplitude $D$ decreases as the noise increases [Fig.~\ref{fig:topo_latent}(f); Confusion matrix shown in SI Fig.~S7(c)].


\subsection*{Cartographer detects nonlinear transients}

We show above that the cartographer segregates systems by topology. However, by construction, $h$ is not fully diffeomorphism-invariant, and it is thus expected to be sensitive to geometric information as well. This suggests that the cartographer should distinguish more subtle forms of global structure beyond topologically invariant features. 

While there exist statistical tests and canonical observables to identify common behaviors \cite{christ_time_2018}, understanding, implementing, and running all possible tests without prior knowledge and assumptions demands considerable researcher and computer time. Our method provides general capabilities that enable quick exploration of dynamical data. We illustrate such a potential use case for the cartographer.

\begin{figure*}
    \centering
    \includegraphics[scale=1]{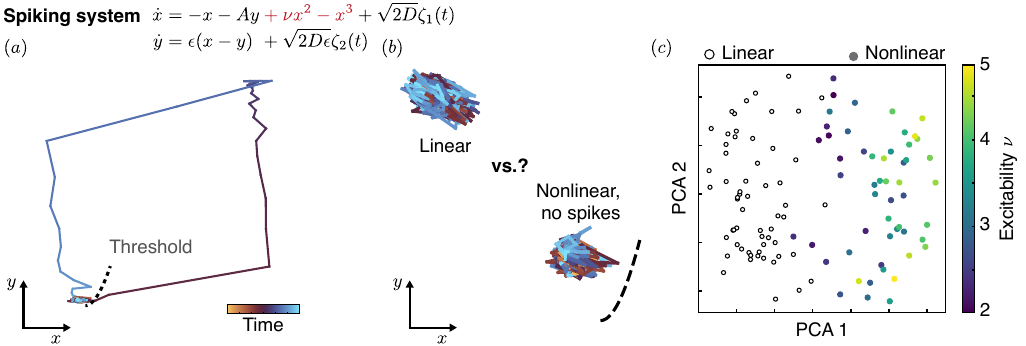}
    \caption{\textbf{The cartographer identifies local nonlinearities.}  \emph{(a)} A noisy excitable system irregularly exhibits spiking trajectories. \emph{(b)} Trajectories of an excitable system that has not spiked (has not crossed the threshold indicated by the dashed line) and its linear approximation which cannot spike are visually similar. \emph{(c)} Embedding trajectories shows that nonlinear systems are discernible from linear systems in the latent space.}
    \label{fig:usecases}
\end{figure*}

The cartographer $h$ provides the ability to compare inputs for which there are no obvious discriminating observables.  Consider the problem of distinguishing between a linear system and a nonlinear excitable system with same linear dynamics which has not spiked. The presence of nonlinearities leads to subtle differences in the trajectory statistics, which reflect the existence of a threshold in phase space: beyond this threshold, trajectories escape the stable domain before coming back from a different direction [Fig.~\ref{fig:usecases}(a)]. We find that our system can distinguish between both populations without prior indications [Fig.~\ref{fig:usecases}(c)]. The map thus provides a way to turn nonlinear analytical problems into standard statistical discrimination tests in latent space, without having to design specific statistical tests, which, for this example, would be based on higher-order moments of the trajectories (SI Sec.~VIII). 

Another application discussed in SI Sec.~IX is the detection of non-potential dynamics, by comparing data generated by systems deriving from a potential $\dot{\mathbf{x}} = - \nabla V(\mathbf{x})$ against observed trajectories (Fig.~S11). As a last example, inspired by the notion of canonical systems used in theoretical neuroscience \cite{izhikevich_neural_2000}, we can also use geometric structure in the trajectory data to introduce a finer categorization of latent space beyond topologically invariant features (SI Sec.~X, Fig.~S12).

\subsection*{Contrastive learning assists experimental analysis}

The contrastive learning strategy allows us to identify meaningful features summarizing complex dynamics. 
However, many dynamics may not be fully reducible to a low-dimensional manifold, and experimental measurements of them are likely to be only partial characterizations of the system state.
Can we still use our self-supervised learning strategy to characterize such data?

\begin{figure*}
    \centering
    \includegraphics[scale=1]{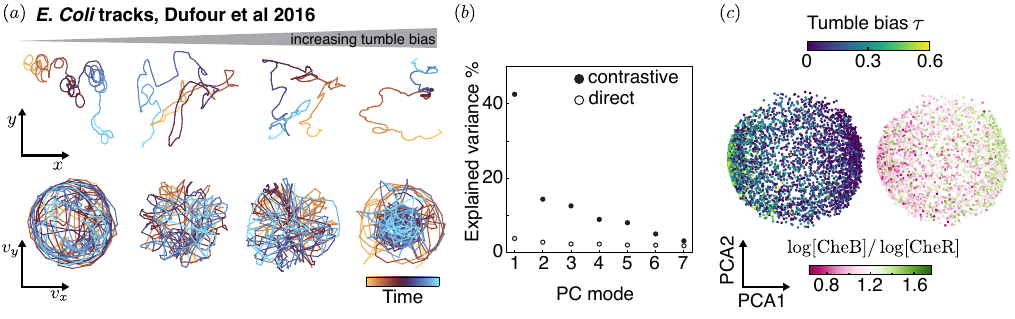}
    \caption{\textbf{Contrastive cartography for bacterial behavior characterization.}  \emph{(a)} Single bacterial tracks alternate between running phases (higher velocity, more persistent orientation) and tumbling phases (slower velocity, rapid orientation change). \emph{(b)} As a representation tool, contrastive cartography produces embeddings with better linear separability than the original data, with more variance captured along the first few principal components. \emph{(c)} Principal component 1 recovers tumble bias as a strong correlate to this behavioral axis, as well as the ratio of the proteins $\log[\text{CheB}]/\log[\text{CheR}]$. (See SI Sec XI, Fig.~S14 for t-SNE representation and additional correlates.)}
    \label{fig:runNtumble}
\end{figure*}

As an example, we consider a publicly available dataset of run-and-tumble motion of \emph{E.\ coli} bacteria \cite{dufour_direct_2016} [Fig.~\ref{fig:runNtumble}(a)]. These dynamics are commonly modeled by a stochastic process with a complex high-dimensional control system, and considerable effort has been devoted to designing statistics to characterize the trajectories. We ask whether the contrastive approach can identify meaningful behavioral variability, without requiring prior knowledge. We expect that important features of the velocity dynamics are invariant under orthogonal transformations of the coordinate system $(x,y)$, and that a contrastive approach may be useful for identifying these invariant features.

The data include measurements tracking wild type \emph{E.\ coli} and mutants which vary with respect to their probability to tumble (tumble bias), which affects chemotaxis \cite{dufour_direct_2016}. 
Here, we train a neural network on the experimental velocity trajectories $\left(v_x(t), v_y(t)\right)$ of the wild type, with the same architecture (SI Sec.~II.A) and InfoNCE loss as the previous cartographer, using data augmentations in the orthogonal group $O(2)$ (SI Sec. XI). We then use the trained cartographer to embed the velocity trajectories of the mutants. 
Evaluating the first principal component of the embedding, we find that it explains a large part of the variance, in contrast to the principal components obtained from a direct analysis of the velocity data [Fig.~\ref{fig:runNtumble}(b)]. This suggests that enforcing rotational invariance improves the ability to identify the key low-dimensional feature of these dynamics. This is also the case when using non-linear dimensional-reduction methods (SI Sec XI, Fig.~S14). 

What is this feature? Examining correlations of experimental measurements with the first principal component, we recover the tumble bias as the most salient feature of the observed dynamics [Fig.~\ref{fig:runNtumble}(c)]. Consistent with the analysis of \cite{dufour_direct_2016}, the first principal component also correlates strongly with the ratio of the logarithms of the concentrations of the proteins $\text{[CheB]}$ and $\text{[CheR]}$, demonstrating a direct link between behavioral features and biochemical parameters. 
Determining the tumble bias from experimental bacterial tracks required the authors of \cite{dufour_direct_2016} to construct a probabilistic model to identify tumbling and running periods. The fact that the symmetry-informed contrastive approach introduced here yields the known relationships without the need for such a system-specific model shows the usefulness of the approach for interpreting experimental data.

\section*{Discussion}


Here, we show how the minimization of a contrastive learning objective, paired with a data-augmentation strategy enforcing coordinate invariance, allows learning an interpretable representation of multidimensional dynamical trajectories even with a relatively simple neural network. 
Our work provides a physics-informed approach for using machine learning with dynamical data distinct from the common generative model, in which machine learning-based regressions are used to find optimal dynamical update rules to extrapolate dynamics from data~\cite{gilpin_generative_2024}. Indeed, having a generative model, whether symbolic or data-driven, does not directly provide a qualitative characterization because analytical methods can rapidly become limited in the face of nonlinearities and growing numbers of parameters.


Here, we focused on 2D dynamical systems. 3D behaviors are even richer, with the possibility of chaotic dynamics and a wider zoology of non-chaotic attractors. Our approach can be used to generate a similar training ensemble and embedding map for higher-dimensional systems, which would be an interesting subject for future work. Beyond 3D, the training ensemble of cubic-order models becomes harder to sample and more refined forms of sampling (e.g., \cite{chi2024sampling}) might become necessary to provide a sufficiently rich training ensemble. The contrastive strategy could also be used to train a cartographer for different families of dynamics, such as different stochastic processes. It is also possible to train the cartographer directly on data, as is done here in Fig.~\ref{fig:runNtumble}. In this case, the interpretability of the latent space is conditioned on identifying which observables correlate most with different latent space regions.


An additional possible application of our method is to leverage the ``phase diagram'' encoded by the latent space to identify tipping points in complex systems, a problem which has attracted both classical and data-driven work \cite{scheffer_earlywarning_2009,dai_generic_2012,bury_deep_2021, huang_deep_2024}. As dynamics in the vicinity of a bifurcation (or generalizations thereof in the presence of noise) are expected to be included in the polynomial class of systems used to train the network, one could use low-dimensional representations of the complex system dynamics to detect sudden changes in embedding without assumptions on the specific kind of bifurcations present in the system.

One could also hope to leverage nonlinear regime cartography as a heuristic to help infer interpretable candidate models from data,  similar to structure or sparsity \cite{brunton_discovering_2016,baddoo_physics-informed_2023}. A heuristic to pick a simple model amenable to theory work---e.g., a normal form that is close in some sense to the observed dynamics---could help provide the elements of a Landau-like phenomenological description. By providing otherwise inaccessible observables characteristic of canonical nonlinear models, such a construction could be relevant even in nonequilibrium and weakly symmetric systems where global phase-space features are often relevant, such as animal behavior \cite{berman_mapping_2014}, biological development, or immunity \cite{lebel_excitable_2023}.

Recent methods for characterizing dynamics for such complex systems involve fitting effective linear time-evolution operators that can then be used as fingerprints of the underlying dynamics \cite{costa_adaptive_2019,sridhar_uncovering_2024,romeo_learning_2021,cohen_schrodinger_2023}. Our trajectory-based work is complementary to those operator-theoretic approaches, trading the inference of linear, often higher-dimensional operators for an interpretable latent space that captures nonlinear effects in low-dimensional systems. Combining the strengths of both approaches could help provide widely applicable characterizations that capture nonlinear phenomena.

As shown in Figure~\ref{fig:topo_latent}, a fundamental limitation is that sufficient amounts of noise obscure features of the underlying fixed-point topology, making classification less reliable. An additional limitation arises from the challenges inherent to characterizing multiscale phenomena, whether in time or in amplitude. In particular, detecting oscillations requires sampling faster than all relevant frequencies. Observing and successfully identifying rare events could also require prohibitively long data inputs which our current architecture might have trouble supporting.

While it is now possible to embed practically any data into a vector space \cite{mehta_theoryisdead_2024}, making such embeddings interpretable and useful for further theory work is an outstanding challenge. Using symmetries and invariances that are common in physics, contrastive learning can identify forms of order that are otherwise difficult to characterize \cite{gokmen_symmetries_2021}. With time-resolved high-dimensional data---whether from biological imaging, ``-omics'' datasets, or large-scale sensor batteries---becoming increasingly common, we hope that this strategy makes identifying distinct behaviors a simpler task on the path to model and control complex systems.

Code is available at \url{https://github.com/NicoRomeo/DynCarto}.

\begin{acknowledgments}
We gratefully acknowledge helpful discussions and code from Arvind Murugan, Matthew S. Schmitt, Peter Y. Lu, Hermann Riecke and Daniel S. Seara. NR acknowledges support from the University of Chicago Biological Physics and Center for Living Systems Fellowships. ERJ acknowledges support from the Burroughs Wellcome Foundation through a Career Award at the Scientific Interface, and the Clare Boothe Luce Foundation (ERJ). This research was supported by the Physics Frontier Center for Living Systems funded by the National Science Foundation (PHY-2317138). Computing resources were provided by the University of Chicago Research Computing Center.
\end{acknowledgments}

\bibliography{biblio}

\end{document}


\preprint{APS/123-QED}

\title{Supplement to: Characterizing nonlinear dynamics by contrastive cartography}

\author{Nicolas Romeo}
\email{nromeo@uchicago.edu}
\affiliation{%
 Center for Living Systems, University of Chicago, Chicago, Illinois 60637
}%
\affiliation{%
 Department of Physics, University of Chicago, Chicago, Illinois 60637
}
\author{Chris Chi}%

\author{Aaron R. Dinner}
\email{dinner@uchicago.edu}
\affiliation{%
 Center for Living Systems, University of Chicago, Chicago, Illinois 60637
}%
\affiliation{%
 Department of Chemistry, University of Chicago, Chicago, Illinois 60637
}%

\author{Elizabeth Jerison}
\email{ejerison@uchicago.edu}
\affiliation{%
 Center for Living Systems, University of Chicago, Chicago, Illinois 60637
}%
\affiliation{%
 Department of Physics, University of Chicago, Chicago, Illinois 60637
}%

\date{\today}%

\maketitle
\small{
\tableofcontents
}

\section{Random dynamical system ensemble}
\label{sec:trainingdata}

Our contrastive learning system is trained using a broad, randomly-generated class of dynamical systems that covers many common models in science and engineering. In this section, we define the two-dimensional dynamical systems used to train our neural networks. This ensemble is also used to explore the geometry of the latent space with respect to the dynamical flow structure (See Sec.~\ref{sec:latent_structure}). 

As smooth functions in the plane can be expanded onto the polynomial basis, we here consider as a generic parametrization of the flow written in components $\mathbf{f} = (\mathrm{f}_x, \mathrm{f}_y)$
\begin{subequations}
\begin{align}
    \mathrm{f}_x & = \sum_{0<i+j\leq 3} a_{ij} x^i y^j \\
    \mathrm{f}_y & = \sum_{0<i+j\leq 3} b_{ij} x^i y^j
\end{align} \label{eq:rand_poly_app}%
\end{subequations}
with real-valued coefficients $a_{ij}$ and $b_{ij}$, denoted in the main text by $c^1_{ij}$ and $c^2_{ij}$ respectively. To create a random ensemble, simply choosing coefficients independently sampled from the unit normal distribution $\mathcal{N}(0,1)$ will usually lead to divergent solutions, which are not representative of typical measurements on controlled natural systems.
To enforce boundedness on the flow, we impose restrictions on the flow away from the origin. A simple criterion to avoid diverging solutions is to prevent the flow from pushing the system away to infinity by imposing $\mathbf{f}\cdot \mathbf{x} \leq 0$ as $||\mathbf{x}||\rightarrow \infty$, which in terms of parameters is
\begin{align}
    a_{30}x^4 + (a_{21}+b_{30}) x^3 y  +(a_{12}+b_{21})x^2 y^2 & \\+ (a_{03}+b_{12}) x y^3 + b_{03}y^4 & \leq 0 
\end{align}
This inequality must hold for any large value of $||\mathbf{x}||^2=x^2+y^2$, and is met under the constraints
\begin{subequations}
\begin{align}
    a_{21}+b_{30} &= 0\\
    a_{03}+b_{12} &= 0\\
    a_{30}x^4 + (a_{12}+b_{21})x^2 y^2 +b_{03}y^4 & \leq 0
\end{align}
\end{subequations}
Solving the biquadratic inequality, we find the equivalent conditions 
\begin{subequations}
    \begin{align}
        a_{21} & = -b_{30} \\
        a_{03} & = - b_{12} \\
        a_{30} & \leq 0\\
        b_{03} & \leq 0\\
        (a_{12}+b_{21})^2 & < 4 |a_{30} b_{03}|  \label{eq:intermediate_stab_condition}
    \end{align} \label{eq:stab_conditions_ensemble}%
\end{subequations}
To simplify our sampling procedure, we tighten condition Eq.~\eqref{eq:intermediate_stab_condition} by imposing the stronger condition $a_{12}+b_{21} =0$.

The resulting ensemble is therefore sampled by sampling all 18 coefficients $a_{ij}, b_{ij}$ from the unit normal distribution $\mathcal{N}(0,1)$, then replacing $a_{30} \leftarrow - |a_{30}|, b_{03} \leftarrow - |b_{03}|$. Finally, we set $b_{30} \leftarrow -a_{21}, b_{12} \leftarrow - a_{03}$, and $b_{21} \leftarrow - a_{12}$. The total number of independent coefficients is thus reduced to 15.

To simulate equations on fixed time- and length-scales, we introduce the time rescaling $t' = t/T$ with $T = 1/|a_{10}|$ and rescaling coordinate axis by $x' = x/X$, $y'=y/Y$ with  $X= \sqrt{-|a_{10}|/a_{30}}$, $Y= \sqrt{-|a_{10}|/b_{03}}$. This choice sets the stabilizing cubic terms to $-1$ in the rescaled dynamical equations. For training, the noise in SDE simulations is rescaled by the scaling factors $X,Y$ leading to an anisotropic additive noise with $D_X = DT/X^2$, $D_Y=DT/Y^2$.
For training the neural network, we use $2\cdot 10^4$ such systems, integrated over the interval $t'\in [0,50]$. The SDEs are integrated in this reduced coordinate system using \texttt{DifferentialEquations.jl} with the high-order adaptive method \texttt{SKenCarp} \cite{rackauckas_diffeq_2017}. 
For each trajectory, we use a randomly-chosen fixed point as initial condition. The effect of changing initial condition on training are further investigated below: the effect of choosing random initial points is detailed in Sec.~\ref{sec:alternative_IC} and Fig.~\ref{fig:alternativeIC}, while the effect of cutting out transient is explored in Sec.~\ref{sec:transient_latent} and Fig.~\ref{fig:comp_viz_transients}.

\section{Contrastive learning}
\label{sec:contrastive_learning}

Contrastive learning is a self-supervised learning framework that leverages the invariance of properties (or, more generally, the invariance of \emph{semantic meaning}) of the data $\mathbf{X}$ to specific transformations $\mathbf{X} \mapsto \tilde{\mathbf{X}}$, termed \emph{augmentations}, to learn a lower-dimensional representation of the data which preserves the desired invariance or meaning \cite{chen_simple_2020}. We use this approach to train an encoder, a function $h$ mapping high-dimensional data input $\mathbf{x}$ to a lower-dimensional latent space representation $h(\mathbf{x})$. In this section, we detail our network architecture, input data format, and augmentation design.

\subsection{Network architecture and loss function}
\label{sec:net_architecture}

The encoder $h$ is realized by a fully connected three-layer feed-forward multilayer perceptron (MLP). The MLP has constant width, with each layer having $n$ neurons with ReLu activation function, and outputs a $n$-dimensional vector. We use the \texttt{InfoNCE} loss function, where NCE stands for `Noise Constrastive Estimation', a common contrastive objective function that has been successfully employed previously used in physical problems and to study dynamical data \cite{oord_representation_2019,murphy_machine-learning_2024, schmitt_information_2024,gokmen_statistical_2021,gokmen_symmetries_2021}.

The \texttt{InfoNCE} loss promotes similarity between the embeddings of the samples and their augmented versions by using a classification loss: Two samples are positively related if they are both the result of augmentations of the same input data, and negatively otherwise. The classification loss is a standard cross-entropy loss that encourages distinction between positive and negative samples.
More formally, a batch of $N_B$ samples is sampled from a precomputed trajectory dataset. Each sample $\mathbf{X}_i$, $i=1,\ldots,N_B$ is associated with an augmented version of itself $\tilde{\mathbf{X}}_i$, and embedded by passing through the encoder to obtain vectors $h(\mathbf{X}_i)$, $h(\tilde{\mathbf{X}}_i)$. We then form the $N_B \times N_B$ similarity matrix $f_{ij}$ whose entries are the cosine similarity between untilded and tilded embeddings 
\begin{equation}
    f_{ij} = h(\mathbf{X}_i) \cdot h(\tilde{\mathbf{X}}_j)
\end{equation}
These $f_{ij}$ are interpreted as log-probabilities of $i$ and $j$ being positively related:
The InfoNCE loss $\ell_\mathrm{InfoNCE}$ is given by the cross-entropy of the probability associated with $f$ and the identity matrix
\begin{equation}
    \ell_\mathrm{InfoNCE} = \frac{1}{N} \sum_{i,j=1}^{N_B} \delta_{ij}\log \frac{\exp{f_{ij}}}{\sum_{k=1}^{N_B} \exp{f_{kj}}}.
\end{equation}
We note that the InfoNCE loss has strong grounding in information theory: minimizing the InfoNCE loss is equivalent to maximizing the mutual information between positive samples \cite{oord_representation_2019, gokmen_symmetries_2021}.

\begin{figure*}
    \centering
    \includegraphics[scale=1]{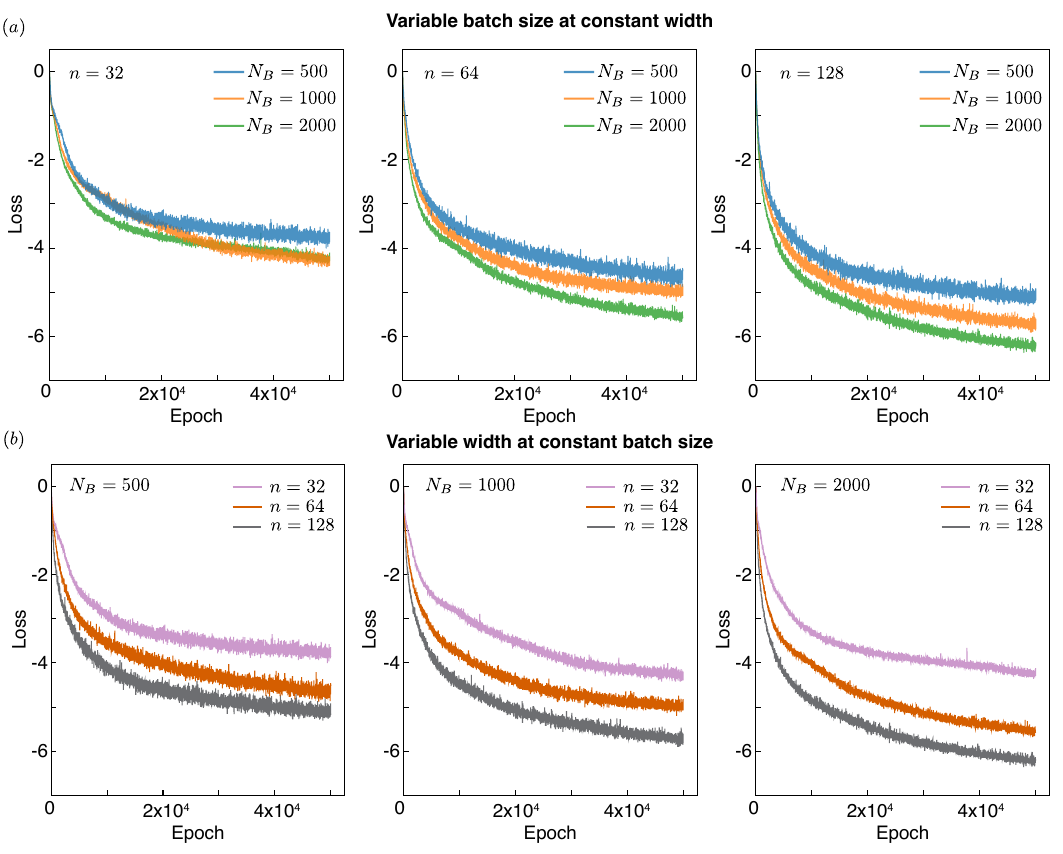}
    \caption{\textbf{Influence of hyperparameters on learning} \emph{(a)} Learning curves for different batch sizes for set network width. Larger batch sizes lead to better learning, as is often expected for constrative losses.  \emph{(b)} Learning curves for different netowrk width for at set batch size. Larger $n$, and thus a wider network and more expressive latent space, leads to better learning outcomes.}
    \label{fig:MLhyper}
\end{figure*}

We use \texttt{pytorch} \cite{paszke_pytorch_2019} to implement the machine learning pipeline. For practical selection of hyperparameters $N_B, n$, we note that contrastive learning objectives often benefit from large batch sizes to provide enough negative examples to `contrast' against \cite{chen_simple_2020} - we experiment with variable batch size and network width $n$ and select $N_B = 2000$, $n=128$ (Fig.~\ref{fig:MLhyper}). Optimization uses the Adam algorithm \cite{kingma_adam_2017} with learning rate $10^{-4}$ and other parameters set to defaults.

We note that our MLP is a very simple network with very little structure (or `inductive biases') imposed internally. Future work could benefit from other neural network architectures, such as variable layer widths or Convolutional Neural Networks (CNN). CNNs in particular often perform better than MLPs on data where sequential information is important, such as the time series considered here.

\subsection{Augmentations}
\label{sec:augmentations}

Here, we discuss the input format of our data and the construction of augmented data to generate positive samples for the contrastive learning objective.
A single input data of our neural network consists of $N$ distinct $d$-dimensional trajectories with $T$ equispaced timepoints represented as a data array $\mathbf{X} = (x^{i,n}_t)$ with $i\in\{1,\ldots, d\}, n\in \{1, \ldots, N\}, t\in\{1,\ldots,T\}$. For every neural network discussed in this paper, $N=30$ and $T=100$.

To generate a random linear invertible transformation, we sample a random $d\times d$ matrix  $R$ with coefficients sampled from the unit normal distribution $\mathcal{N}(0,1)$, and compute its Singular Value Decomposition such that 
\begin{equation}
    R = U \Sigma V^T,
\end{equation}
with $U, V$ $d\times d$ orthogonal matrices (satisfying $U^TU = V^TV = I$) and $\Sigma$ diagonal. The matrix $Q=UV^T$, equivalent to the orthogonal matrix obtained from the $QR$-decomposition, now uniformly samples the orthogonal group $O(d)$ that generates rigid rotations and reflections \cite{mezzadri_howto_2007}. To allow for shearing of the coordinate axes, we additionally sample a random matrix
\begin{equation}
    S = \begin{pmatrix}
        1+s_{00} & s_{01}/r & s_{02}/r & \cdots & s_{0d}/r \\  s_{10}/r & 1+s_{11} &  s_{12}/r & \cdots & s_{1d}/r\\
        s_{20}/r & s_{21}/r & 1+s_{22} & s_{23}/r & \cdots \\
        \ddots & \ddots & \ddots & \ddots & \ddots
    \end{pmatrix}
\end{equation}
where $s_{ij}$ are all independently sampled from $\mathcal{N}(0,1)$ and $r$ is a scaling factor to constrain the shearing and avoid numerical ill-conditioning.
In $d=2$, we use specifically
\begin{equation}
    S = \begin{pmatrix}
        1+s_{00} & s_{01}/r  \\  s_{10}/r & 1+s_{11}
        \end{pmatrix}
\end{equation}
with the choice of $r=5$.
The resulting linear transformation 
\begin{equation}
    M = U S V^T
\end{equation}
can be intuitively understood as applying the shearing matrix $S$ in the coordinate axis defined by $V^T$, then pulling back the resulting object to the coordinate system defined (in the original coordinates) by $UV^T$.

The augmented samples $\tilde{\mathbf{X}}$ are then given by the data array
\begin{align}
\tilde{x}^{i,n}_t = \sum_{j,k} \sigma^n_k M_{ij} x^{j,k}_t 
\end{align}
with the matrix $\sigma^n_k$ representing the action of a random permutation on the order of trajectories (in the group $S_N$) to softly enforce the invariance of the neural network output on the order of input trajectories.

Finally, before being fed to the neural network, all data arrays $X$ are centered and normalized by subtracting their mean and dividing element-wise by their variance.

We note that this choice of augmentation, with $UV^T \in O(d)$ and no restriction on the sign of the determinant of $S$, allows for coordinate transformations with negative determinants. Such transformations do not preserve orientation: With this choice of augmentation, it is thus not possible to distinguish, for instance, left- and right-handed limit cycles. Imposing an additional step if $\mathrm{det}M <0$, in which we multiply $M$ with a negative-determinant matrix such as $P = \mathrm{diag}(1, -1)$ would enforce orientation-preserving augmentations. 

\section{Phase identification in low-dimensional systems}

In this section we present the different dynamical systems used to test the cartographer. With the exception of the later analysis using different initial conditions for the training data, all tests are done using the same pre-trained neural network using data generated from the protocol in Sec.~\ref{sec:trainingdata}.
All numerical integration of SDEs is done using \texttt{DifferentialEquations.jl} with the high-order adaptive method \texttt{SKenCarp} \cite{rackauckas_diffeq_2017}. 

\subsection{FitzHugh-Nagumo}
\label{sec:fhn}

Our first test case in the main text is the identification of distinct behaviors in FitzHugh-Nagumo systems with varying parameters, for which the flow is given by \cite{fitzhugh_impulses_1961,Nagumo_active_1962,izhikevich_dynamical_2006}
\begin{subequations}
    \begin{align}
        \mathrm{f}_x & = x - \frac{x^3}{3} - y + I \\ 
        \mathrm{f}_y & =  \epsilon(x + a - by)
    \end{align} \label{eq:FhN}%
\end{subequations}
Depending on the parameters $(a,b, \epsilon, I)$, the system can be monostable, bistable, excitable or it can oscillate. 

For Main text Fig.~1(b), we solve the SDEs
\begin{equation}
    \mathrm{d}\mathbf{x} = \mathbf{f}(\mathbf{x})\mathrm{d}t + \sqrt{2D} \mathrm{d}\boldsymbol{\zeta}
\end{equation}
with the following mean parameter values:
\begin{itemize}
    \item \textbf{Monostable:}  $a= 0.2, b=0.5, \epsilon =100/12.5, I =0.5$
    \item \textbf{Oscillator:} $a= 0.2, b=0.5, \epsilon = 1.0, I =0.5$
    \item \textbf{Bistable:} $a=1 , b=2.0, \epsilon =1/12.5, I =0.5$
    \item \textbf{Excitable:} $a= 0.8, b=1.0, \epsilon =1/12.5, I = 0.1$. We note that in this case, the flow only has one fixed point - Monostable and Excitable systems are topologically equivalent.
\end{itemize}
For all systems in Main text Fig.~1(b), we pick $D=0.1$ as noise amplitude and we let each parameter $\mu \in \{a,b,\epsilon,I\}$ vary around its mean value $\bar{\mu}$ given above by sampling its actual value $\mu$ from the normal distribution $N(\bar{\mu}, 0.05\bar{\mu})$. After embedding trajectories from $40$ distinct parameter groups of each category, we see in a 2-dimensional Multidimensional Scaling projection (MDS) the emergence of clusters reflecting the distinct behaviors.

\subsection{SNIC-Hopf}
\label{sec:SNICHopf}

As an analytically tractable dynamical system exhibiting distinct bifurcations, we use the SNIC-Hopf toy model studied in \cite{martin_transition_2024} to test our dynamical characterization approach. The model is given by 
\begin{equation}
    \frac{\mathrm{d}}{\mathrm{d}t}\begin{pmatrix}x \\ y \end{pmatrix} = \begin{pmatrix}
        k & \alpha_+-\alpha_- \\ \alpha_+ +\alpha_- & k
    \end{pmatrix} \begin{pmatrix}
        x \\ y 
    \end{pmatrix} - (x^2+y^2) \begin{pmatrix}
        x \\ y 
    \end{pmatrix} \equiv \mathbf{f}_\text{SNIC/Hopf}(\mathbf{x}). \label{eq:snichopf_eq}
\end{equation}
For convenience, we rederive here the mean-field phase diagram for the case $\alpha_+=1$ presented in the main text, and refer to \cite{martin_transition_2024} for more general results.

We first remark that the origin $(0,0)$ is a fixed point of Eq.~\eqref{eq:snichopf_eq}.
The Jacobian of the system at $(0,0)$ is given by
\begin{equation}
    J_{0} = \begin{pmatrix}
        k & 1-\alpha \\ 1+\alpha & k
    \end{pmatrix}
\end{equation}
which has trace $2k$ and determinant $k^2 + \alpha^2 - 1$. If $k>0$, the origin is always unstable. If $k<0$, the determinant is positive if $k^2 + \alpha^2>1$, which is true outside of the circle of radius 1 around the origin in the $(k,\alpha)$ plane. Thus, the origin is stable for parameter values outside the unit circle for negative $k$, or put differently if $k<0$ and $k^2>1-\alpha^2$.

\begin{figure}
    \centering
    \includegraphics[width=0.5\linewidth]{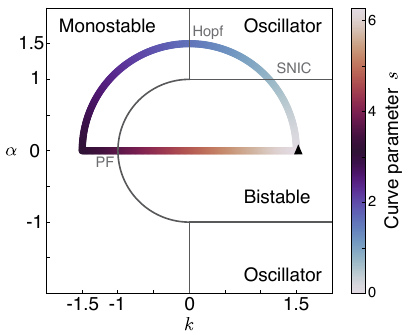}
    \caption{\textbf{Path in parameter space to probe the effect of bifurcations in latent space} Path $\left(k(s), \alpha(s)\right)$ sampled at 500 points. For $s\in [0, \pi], k(s) = 1.5\cos(s)$ and $\alpha(s) = 1.5\sin(s)$. For $s\in[\pi, 2\pi], k(s) = 3(s/\pi- 3/2)$, $\alpha(s) =0$. }
    \label{fig:snic_hopf_path}
\end{figure}

Non-zero fixed points have norm-squared $\Lambda^2$ satisfying the eigenvalue equation
\begin{equation}
    \begin{pmatrix}
        k & 1-\alpha_- \\ 1 +\alpha_- & k
    \end{pmatrix} \begin{pmatrix}
        x \\ y 
    \end{pmatrix} = \Lambda^2 \begin{pmatrix}
        x \\ y 
    \end{pmatrix} \label{eq:eigenproblem_snic}
\end{equation}
which allows solutions satisfying
\begin{equation}
    (\Lambda^2-k)^2 = 1-\alpha^2.
\end{equation}
If $|\alpha| >1$, then there are no other solutions than the origin. If $|\alpha| <1$, then there are other, non-zero fixed points if
\begin{equation}
    \Lambda^2_\pm = k \pm \sqrt{1-\alpha^2} > 0
\end{equation}
Finding the eigenvectors of norm $\Lambda^2_\pm$ of Eq.~\eqref{eq:eigenproblem_snic} when they exist, we have
\begin{align}
    \mathbf{x}^\sigma_{\pm} = \sigma\sqrt{\frac{\Lambda^2_\pm}{2}}\begin{pmatrix}
        \sqrt{1- \alpha^2} \\ \pm\sqrt{1+ \alpha^2} 
    \end{pmatrix}
\end{align}
with $\sigma = \pm$, giving up to 4 solutions. If $k^2 + \alpha^2 <1$, only the two solutions associated with $\Lambda_+$ exist and these are stable. If $k^2 + \alpha^2 >1$, the solutions corresponding to $\Lambda_-$ give two additional unstable fixed points.

When $k>0$ and $|\alpha|>1$, since the system is asymptotically stable and there are no other fixed points than the unstable origin, the Poincare-Bendixson theorem proves that the system has a stable limit cycle. If $k<0$, $k^2+\alpha^2 >1$, the determinant is positive and the system has a single fixed point at the origin. 

To summarize, we obtain 3 distinct dynamical phases:
\begin{itemize}
    \item if $k>0$ and $|\alpha| > 1$, the system has a limit cycle
    \item if $k<0, k^2 > 1-\alpha^2$ then the system has a single stable fixed point at the origin.
    \item $k>-\sqrt{1-\alpha^2}$ when $\alpha^2 < 1$ then the system has two stable fixed points at $\mathbf{x}^+_+$ and $\mathbf{x}^-_+$.
\end{itemize}
Monostable and bistable phases are separated by a Pitchfork (PF) bifurcation, monostable and oscillatory phases are separated by a supercritical Hopf bifurcation as the eigenvalues of the Jacobian at the origin turn complex, and the bistable to oscillatory transition is an example of a (double) Saddle-Node on Invariant Circle (SNIC) bifurcation, with 2 saddle-node pairs appearing on the limit cycle. This is easiest to see in polar coordinates, in which the system can be written as
\begin{subequations}
    \begin{align}
        \dot{r} &= r(k+\sin(2\theta))-r^3 \\
        \dot{\theta} &= \alpha +\cos(2\theta)
    \end{align}
\end{subequations}
The additional factor of $2$ in the trigonometric function compared to the standard SNIC form \cite{strogatz_nonlinear_2024} (see also Sec.~\ref{sec:refsystems}) reflects the presence of an additional $\mathbb{Z}_2$ symmetry in the system leading to the appearance of a second saddle-node pair on the limit cycle defined by $\dot{r}=0$.

\subsection{Saddle-Homoclinic Orbit excitable system}
\label{sec:SHOsystem}

As an additional, more complex example of a parameterized system undergoing both local and global bifurcations, we consider the following system
\begin{subequations}
    \begin{align}
        \dot{x}& = -x +\nu x^2 - x^3 - Ay, \\
        \dot{y}& = \epsilon\left( \frac{x^2}{2} - \frac{x}{2} - y\right).
    \end{align}
\end{subequations}
For specific values of $\nu, A, \epsilon >0$, this system exhibits a Saddle-Homoclinic Orbit  (SHO) bifurcation \cite{izhikevich_neural_2000}, a global bifurcation in which a 
limit cycle disappears as it collides with a saddle point, forming an homoclinic orbit at the critical point. 

Its nullclines are given by
\begin{subequations}
\begin{align}
    y = & (-x + \nu x^2 -x^3)/A \\
    y = & \frac{x^2}{2} -\frac{x}{2}
\end{align}
\end{subequations}
which intersect at $(0,0)$ for all parameter values. For $\nu > 2$,  there are two other fixed points at $x_\pm=\nu/2-A/4 \pm \sqrt{\left(A/2-\nu\right)^2/4 + A/2 - 1}$, and $y_\pm = (x^2_\pm -x_\pm)/2$. 

The fixed point at $(x_-, y_-)$  is stable and the origin is unstable for $A>2$, and $x_- <0$;
A transcritical bifurcation at $A=2$ leads to a loss of stability for the fixed point at $x_-$, which is located right of the origin $x_- >0$ for $A <2$, and correspondingly the origin becomes stable.

In what follows, we focus on the case $A>2$, for which $x_- <0 < x_+$, the fixed point at $x_-$ is always stable, and the fixed point at the origin is an unstable saddle. 

For $A>2$, the right fixed point at $x=x_+$ is a focus, with imaginary eigenvalues. It loses stability when $\epsilon < \epsilon_\mathrm{crit}(A)$ through a supercritical Andronov-Hopf bifurcation, with the critical line given by
\begin{equation}
    \epsilon_\mathrm{crit}(A) = -1 + 2\nu x_+(A) - 3 x_+(A) \label{eq:hopf_SHO}
\end{equation}
with $x_+(A) = \frac{\nu}{2}-\frac{A}{4} + \frac{1}{2}\sqrt{\left(\frac{A}{2}-\nu\right)^2 + 2A - 4}$ as above.

\begin{figure}
    \centering
    \includegraphics[scale=1]{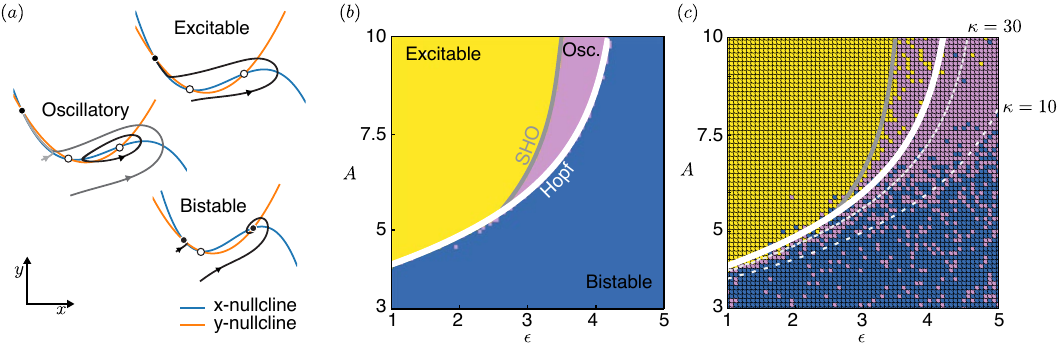}
    \caption{\textbf{Phase identification in Saddle-Homoclinic Orbit system} \emph{(a)} Representative topological structures in each phase. The system exhibits more complex behaviors than the SNIC-Hopf system \emph{(b)} Phase diagram of our minimal model for $\nu=4$. The Hopf bifurcation line is given by Eq.~\ref{eq:hopf_SHO}. The SHO bifurcation line and coloring are obtained using \texttt{Attractors.jl}~\cite{datseris_framework_2023}. \emph{(c)} Inferred phases by k-means clustering in latent space. $\kappa = |\mathrm{Im}\lambda/\mathrm{Re}\lambda|$ is the ratio of decay timescale to oscillation pseudo-period. In the presence of noise, long persistent oscillations lead the system to confuse limit cycles and noisy oscillations about a stable focus.}
    \label{fig:sho_phases}
\end{figure}

As alluded above, the limit cycle arising from the supercritical Hopf bifurcation can disappear when lowering $\epsilon$ or increasing $A$ by growing into the saddle at the origin, providing an example of a SHO bifurcation. We numerically determine the presence of the limit cycle for $\nu=4$ using \texttt{Attractors.jl} \cite{datseris_framework_2023}; a sketch of the nullclines for representative examples of each phase and the numerically-determined phase diagram are shown in Fig.~\ref{fig:sho_phases}(a-b).

We integrate the corresponding SDEs for $D=10^{-4}, \nu=4$ up to $t=50$, and feed the resulting trajectories to the contrastive cartographer. After $k$-means clustering with $k=3$ (results are similar using Agglomerative clustering with Ward linkage), we find that the resulting dynamical classifier recovers mean-field predictions for both the SHO transition between oscillatory and excitable regions, and the Hopf transitions between bistable and monostable (excitable) regions. However, the cartographer has a harder time discovering the expected mean-field separation between the phase with coexisting monostable and limit cycle and the bistable regime (Fig.~\ref{fig:sho_phases}(c)). This can be understood by considering the `quality factor' of the oscillations about the right-most fixed point undergoing the Hopf bifurcation $\kappa = |\mathrm{Im}\lambda| / |\mathrm{Re}\lambda|$, where $\lambda$ is an eigenvalue of the Jacobian at $(x_+, y_+)$. We find that $\kappa$ is very large in the vicinity of the Hopf bifurcation, representative of the phenomenon of critical slowing down. On the finite integration timescales we consider and in the presence of noise, the resulting long-lived noise-induced oscillations induce confusion with the actual limit cycle.

\subsection{Non-linear cell-cycle model}
\label{sec:cellcycle}

We consider here the application of our analysis tool to characterize trajectories generated by a noisy variant of the biochemical model of the cell cycle presented in \cite{tyson_modeling_1991}. The model, summarized by the diagram in main text Fig.~2(d), considers the dynamics of cyclin and cdc2 and their complexes at various stages of phosphorylation.
In line with the notation of \cite{tyson_modeling_1991}, we use the shorthands
\begin{itemize}
    \item[-] $[\mathrm{M}]$ for P-cyclin-cdc2, the maturation promoting factor (MPF) \vspace{-4pt}
    \item[-] $[\mathrm{Y}]$ for cyclin\vspace{-4pt}
    \item[-] $[\mathrm{YP}]$ for cyclin-P\vspace{-4pt}
    \item[-] $[\mathrm{pM}]$ for P-cyclin-cdc2-P (pre-MFP)\vspace{-4pt}
    \item[-] $[\mathrm{C2}]$ for cdc2\vspace{-4pt}
    \item[-] $[\mathrm{CP}]$ for cdc2-P\vspace{-4pt}
\end{itemize}
Additionally, we define the total concentration of cdc2 $[\mathrm{CT}] = [\mathrm{C2}] + [\mathrm{CP}] + [\mathrm{M}] + [\mathrm{pM}]$, the amino acid concentration $[\mathrm{aa}]$ (assumed constant) for the assembly of cyclin, and the ATP concentration $[\mathrm{\sim P}]$ (assumed constant). With these notations, the kinetic equations of the model are given by
\begin{subequations}
    \begin{align}
        \frac{\mathrm{d}[\mathrm{C2}]}{\mathrm{d}t} &= k_6[\mathrm{M}] - k_8[\mathrm{\sim P}] [\mathrm{C2}] + k_9 [\mathrm{CP}]\\
        \frac{\mathrm{d}[\mathrm{CP}]}{\mathrm{d}t} &= -k_3 [\mathrm{CP}][\mathrm{Y}]+k_8[\mathrm{\sim P}] [\mathrm{C2}] - k_9 [\mathrm{CP}]\\
        \frac{\mathrm{d}[\mathrm{pM}]}{\mathrm{d}t} &= k_3 [\mathrm{CP}][\mathrm{Y}] -[\mathrm{pM}] \left(k_4'+k_4\left(\frac{[\mathrm{M}]}{[\mathrm{CT}]}\right)^2 \right) + k_5[\mathrm{\sim P}][\mathrm{M}]\\
        \frac{\mathrm{d}[\mathrm{M}]}{\mathrm{d}t} &= [\mathrm{pM}] \left(k_4'+k_4\left(\frac{[\mathrm{M}]}{[\mathrm{CT}]}\right)^2 \right) -k_5[\mathrm{\sim P}][\mathrm{M}]-k_6[\mathrm{M}] \\
        \frac{\mathrm{d}[\mathrm{Y}]}{\mathrm{d}t} &= k_1[aa] -k_2[\mathrm{Y}]- k_3 [\mathrm{CP}][\mathrm{Y}] \\
        \frac{\mathrm{d}[\mathrm{YP}]}{\mathrm{d}t} &=k_6[\mathrm{M}] - k_7 [\mathrm{YP}]
    \end{align} \label{eq:tysonmodel}%
\end{subequations}
The total cdc2 concentration $[\mathrm{CT}]$ is constant. $k_4$ and $k_6$ are our variable parameters, and the other parameters are given as in \cite{tyson_modeling_1991} by $k_1[\mathrm{aa}] = 0.015$ $\mathrm{min}^{-1}$, $k_2=0$, $k_3[\mathrm{CT}]= 200$ $\mathrm{min}^{-1}$, $k_4' = 0.018$  $\mathrm{min}^{-1}$, $k_5[\mathrm{\sim P}] =0, k_7 = 0.6$  $\mathrm{min}^{-1}$, and $k_8[\mathrm{\sim P}] =10^3 \, \mathrm{min}^{-1}\gg k_9 = 10^2\,\mathrm{min}^{-1} \gg k_6$.

Main text Fig.~2e shows timecourses of $[\mathrm{M}]$ and the total cyclin $[\mathrm{YT}] = [\mathrm{Y}]+[\mathrm{YP}]+[\mathrm{M}]+[\mathrm{pM}]$, with additional weak dynamical (additive) Gaussian noise in all equations in Eqs.~\eqref{eq:tysonmodel} with amplitude $D=0.01$. Input data to the neural network uses the same noise, integrated over $t\in [0, T_f]$, with $T_f = 200$ $\text{min}$. After an initial period of $T_t = 22$ $\text{min}$, the data is sampled at intervals $\Delta t = (T_f-T_t)/100$. The resulting time series are then fed to the cartographer.

For the deterministic dynamics, assuming that $[\mathrm{C2}]/[\mathrm{CT}] \ll 1$ which holds when $k_6 \ll k_9 \ll k_8[\mathrm{\sim P}]$, analytical considerations detailed in \cite{tyson_modeling_1991} give that the high-M steady state is valid when
\begin{equation}
    \frac{k_1[\mathrm{aa}]}{k_6[\mathrm{CT}]} > \sqrt{\frac{k_6}{k_4}}.
\end{equation}
The stable limit cycle exists when
\begin{equation}
   \sqrt{\frac{k_4'}{k_4}}< \frac{k_1[\mathrm{aa}]}{k_6[\mathrm{CT}]} < \sqrt{\frac{k_6}{k_4}},
\end{equation}
and the low-M steady state exists for
\begin{equation}
    \frac{k_1[\mathrm{aa}]}{k_6[\mathrm{CT}]}< \sqrt{\frac{k_4'}{k_4}}.
\end{equation}
Additionally, it was observed in \cite{tyson_modeling_1991} that excitable dynamics are visible in the low-M steady regime in the vicinity of the steady/oscillating transition. Agglomerative clustering in latent space indeed recovers those different regions (Main text Fig.~2(f)).

\subsection{Flow past a cylinder and analysis}\label{sec:flowcylinder}

To show the applicability of our method to data generated by higher-dimensional dynamical processes, we consider the problem of planar hydrodynamic flow past a cylinder in the channel geometry of the DFG 2D-3 benchmark \cite{john_higherorder_2001}.

\subsubsection{Problem definition and numerical resolution}
For a fluid of dynamic viscosity $\eta$,  in dimensionless units where the intake velocity $U$, obstacle diameter $L=5\,\mathrm{cm}$ and density $\rho$ are set to 1, the flow obeys the incompressible Navier-Stokes equations in 2D
\begin{subequations}
    \begin{align}
        \partial_t \bar{\mathbf{u}} + \bar{\mathbf{u}}\cdot \nabla\bar{\mathbf{u}} = & - \nabla \bar{p} + \frac{1}{\mathrm{Re}} \Delta \bar{\mathbf{u}}  \\
        \nabla\cdot \bar{\mathbf{u}} = & 0
    \end{align}
\end{subequations}
where $\bar{\mathbf{u}}$ is the flow and $\bar{p}$ the pressure in dimensionless units, and the Reynolds number $\mathrm{Re}$ is related to the physical values of $U, \rho, L$ and $\eta$ by $\mathrm{Re} = \eta UL/\rho$. Additionally, the fluid obeys no-slip $\mathbf{u}=0$ boundary conditions at the channel edges and on the obstacle, while the intake flow is time- and space-dependent
\begin{equation}
    \mathbf{u}(x=0, y, t) = \left\{\begin{matrix}
        4U\sin(\pi t/2)\frac{y}{H}(1-\frac{y}{H})  \text{ if } t<1\\
        4U\frac{y}{H}(1-\frac{y}{H}) \text{ if } t>1
    \end{matrix}\right.
\end{equation}
in physical units, with $H=0.41\;\mathrm{m}$ the channel height. Finally, the pressure satisfies a Dirichlet condition at the outlet $p(x=L,y,t) =0$.

To numerically solve the problem, we use the reference implementation of the FeniCSx tutorial \cite{baratta_dolfinx_2023,alnaes_ufl_2014,scroggs_basix_2022,scroggs_construction_2022} which can be found at \url{https://jsdokken.com/dolfinx-tutorial/chapter2/ns_code2.html}. Briefly, this implementation integrates the equations on the variable-size triangular mesh defined in the benchmark using a Crank-Nicholson discretization in time. The non-linear term is approximated using a semi-implicit Adams-Bashforth scheme, and the results are validated against the DFG 2D-3 benchmark for the corresponding driving.

\subsubsection{Dynamic Mode Decomposition}

To obtain a low-dimensional representation of the unsteady flow, we compute the Dynamic Mode Decomposition (DMD) of the flow at $\mathrm{Re}=100$ and project flow data for variable $\mathrm{Re}$. 

To construct the DMD basis, we use the SVD formulation exposed in \cite{schmid_dynamic_2010}. At each regularly-spaced timepoint $t\in\{1, \ldots, T\}$ we have our data vector $\mathbf{x}_t = (u^x_1,  \ldots, u^x_N, u^y_1, \ldots, u^y_N)^T$ where $u^x_i$ and $u^y_i$ are the values of the $x$- and $y$-components of the flow at the $i$-th spatial discretization point, $i\in \{1,\ldots,N\}$. We then construct the $2N\times T$ data matrices
\begin{equation}
    X_t = \begin{pmatrix}
        \kern.2em\vline & \kern.2em\vline & & \kern.2em\vline \\
        \mathbf{x}_1 & \mathbf{x}_2 & \cdots & \mathbf{x}_{T-1} \\
        \kern.2em\vline & \kern.2em\vline & & \kern.2em\vline
    \end{pmatrix}, \quad X_{t+1} = \begin{pmatrix}
        \kern.2em\vline & \kern.2em\vline & & \kern.2em\vline \\
        \mathbf{x}_2 & \mathbf{x}_3 & \cdots & \mathbf{x}_{T} \\
        \kern.2em\vline & \kern.2em\vline & & \kern.2em\vline
    \end{pmatrix}
\end{equation}
and take the Singular Value Decomposition of $X_t^T = U S W^T$, where $U$ is a $T\times T$ orthogonal matrix, $S$ is a diagonal $T\times2N$ rectangular matrix, and $W$ is a $2N \times 2N$ orthogonal matrix. We then construct the matrix $\tilde{S} = U^T X_{t+1} W^T S^{-1}$: the DMD basis $\mathsf{w}_k$ are the (complex) eigenvectors of $\tilde{S}$, sorted by the amplitude of their eigenvalues.

\begin{figure}
    \centering
    \includegraphics[scale=1]{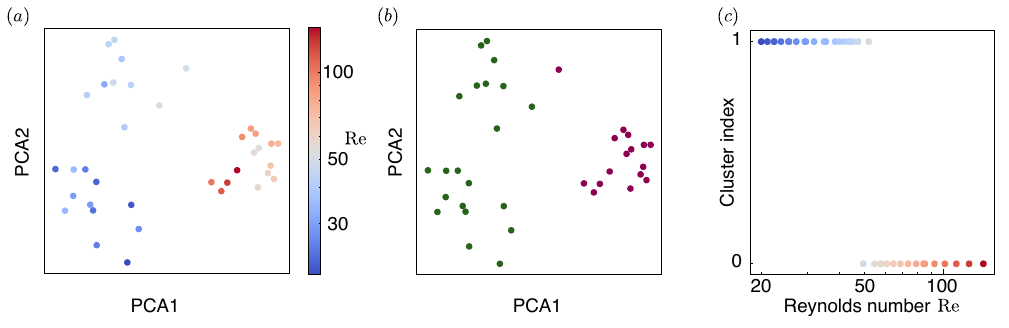}
    \caption{\textbf{Identification of critical Reynolds number by latent space clustering.} \emph{(a)} Visualization of the embedding in PCA coordinates  \textbf{b-c.} $K$-means with $K=2$ identifies clusters aligned with the expectation that $\mathrm{Re}=50$ is the critical Reynolds number. }
    \label{fig:dmd_clustering}
\end{figure}

To project the data onto a 2-dimensional space, we take the (complex) eigenvector $\mathsf{w}_1$ associated with the largest eigenvalue of $\tilde{S}$ of the flow with $\mathrm{Re}=100$, and project flows in this shared basis by computing the matrix product $a = X_t \mathsf{w}_1$. To obtain ensembles of trajectories for each value of $\mathrm{Re}$, we add artificial Gaussian noise of mean zero and standard deviation $\sqrt{2D}\max_t{||\mathbf{x}_t||_\infty}$ with $D=0.05$ to the simulation data.

The resulting data after embedding separates into two clusters aligned with our expectations of steady flow for $\mathrm{Re}\leq 50$ and oscillatory flow for $\mathrm{Re}> 50$ (Fig.~\ref{fig:dmd_clustering}, Main text Fig.~2(ghi)).

\section{Effect of noise on dynamical characterization}
\label{sec:noise_effects}

In Sec.~\ref{sec:SHOsystem} we found that long-lived noise-induced oscillations could cause the neural network to confuse stable fixed points with limit cycles. In this section we examine more closely the influence of noise in the characterization, focusing on a well-characterized example, the pitchfork bifurcation.

\begin{figure*}
    \centering
    \includegraphics[width=\linewidth]{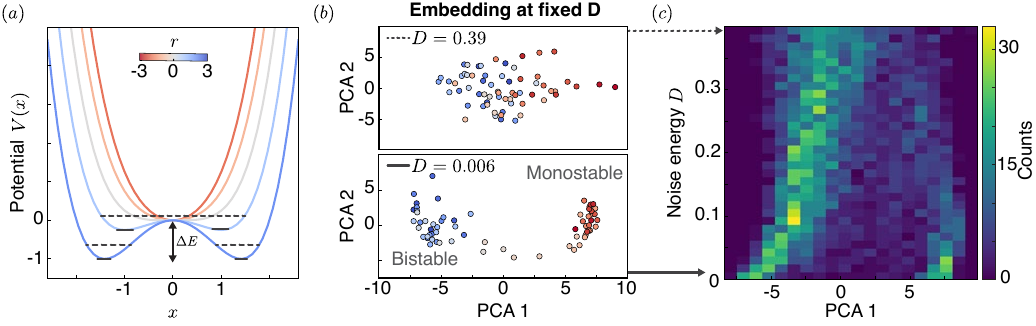}
    \caption{\textbf{Noise blurs the equivalence between flow topology and behavior.} \emph{(a)} Example potential energy landscapes for the $x$-dynamics. For negative $r$, the potential has a single stable fixed point at $x=0$. For positive $r$, there are two stable fixed points at $x = \pm\sqrt{r}$. Low noise levels are indicated by continuous bars, while high noise level by dashed lines. High noise can overcome shallow energy barriers. \emph{(b)} For low noise, the two distinct phases for varying $r$ can be identified in the contrastive embedding space (bottom), while the difference is blurred at high noise (top) (Same panel as Fig.~3(e) of main text). \emph{(c)} As we increase noise level for a set range of $r\in [-3, 3]$, the distinction between embedded clusters is blurred.}
    \label{fig:phasediagram_varnoise}
\end{figure*}

We consider a simple bistable system whose flow derives from a double-well potential through $\mathbf{f}(\mathbf{x}) = - \nabla V(\mathbf{x})$, and
\begin{equation}
    V(\mathbf{x}) = -\frac{r}{2}x^2 + \frac{1}{4}x^4  + \frac{1}{2}y^2
\end{equation}
This potential has a single stable fixed point at $(x,y)=(0,0)$ for $r<0$, and it has two stable fixed points at $(x,y)=(\pm\sqrt{r},0)$ for $r>0$ (Fig.~\ref{fig:phasediagram_varnoise}(a)). The presence of increasing noise in this system is known to progressively blur the boundary between the mono- and bistable regimes \cite{gaspard_spectral_1995, schmitt_information_2024}. In the bistable case, the two stable fixed points are separated by an energy barrier of height $\Delta E = r^2/4$, and the system is expected to spend an average time of order $e^{-D/\Delta E}$ in each well.
For small noise, this timescale can be much longer than the observation time $T_\text{obs}$. As we vary $r$, the contrastive embedding of observed trajectories reveals in principal component coordinates two distinct dynamical phases corresponding to mono- and bistable regimes (Fig.~\ref{fig:phasediagram_varnoise}(b)). 
At high noise the systems frequently jump between wells, blurring the visibility of the two fixed points. Consequently, the two clusters merge as the noise amplitude is increased  (Fig.~\ref{fig:phasediagram_varnoise}(c)).

\section{Construction of dynamical phase diagrams in latent space}
\label{sec:latent_structure}

In this section, we detail the construction and analysis of behavioral regions in the broader latent space spanned by $h$. Essentially, we supplement the dynamical systems used for unsupervised training of the neural network with their numerically-determined fixed-point structure, and use those to train a supervised classifier in latent space. We integrate systems sampled from the ensemble specified in Sec.~\ref{sec:trainingdata}, with the difference that the noise amplitude $D$ is now chosen such that $DT = \text{constant}$ with $T=1/|a_{10}|$. This choice maintains noise levels commensurate with the strength of the linear response.

\subsection{Numerical determination of fixed-point structure}

To obtain a ground truth topological structure of the many dynamical systems we sample and integrate, we numerically determine the location of their fixed points by Homotopy Continuation (HC). We use the \texttt{HomotopyContinuation.jl} solver using the polyhedral starting strategy, automatic differentiation, a maximum of 2000 endgame steps and 10000 tracking steps \cite{breiding_homotopy_2018}. 
We then evaluate the analytical Jacobian obtained directly by differentiating the polynomial expressions at each fixed point determined by HC. We then determine the linear stability of the fixed points by computing the eigenvectors of the Jacobians.

\subsection{Fixed-point statistics}
\label{sec:fp_stats}

\begin{figure*}
    \centering
    \includegraphics[scale=1]{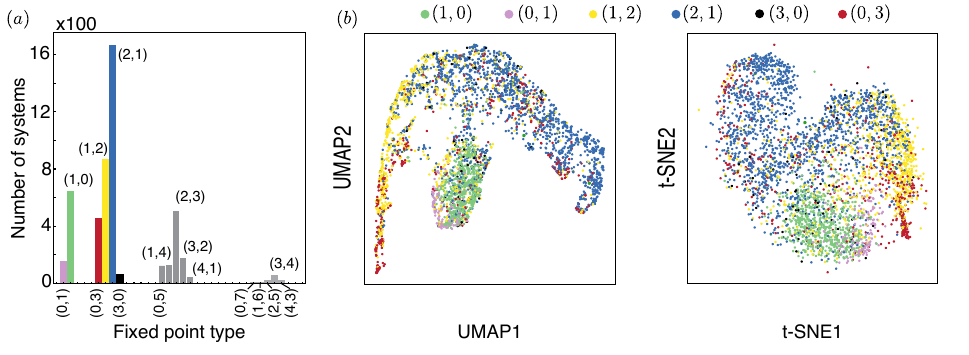}
    \caption{\textbf{Higher-order fixed point structures} \emph{(a)} Number of systems with a given number of stable, unstable fixed points from a sample of $5000$ systems.  \emph{(b)} t-SNE projection of the training data labeled by number of stable and unstable fixed points $(n_\mathrm{stab}, n_\mathrm{unstab})$, restricted to samples with no more than $3$ fixed points.}
    \label{fig:higher_fps}
\end{figure*}

Cubic order systems of two equations in two variables can have up to nine fixed points. We find that about half of our sampled systems have up to $3$ fixed points (Fig.~\ref{fig:higher_fps}(a)), and thus we focus on these simpler topologies which are better sampled. 
Given that we focus on bounded flows in the plane, our systems have topological index $1$ \cite{strogatz_nonlinear_2024} --- to maintain the index, a system with only one unstable fixed point must have a stable limit cycle to compensate. Generally, the index restriction only allows for systems with an odd number of fixed points, and certain combination (e.g. $3$ stable fixed points but no unstable fixed points, or $3$ unstable fixed points but no stable fixed points) are only possible if we account for the existence of stable or unstable limit cycles.

A full description of the topological structure of two-dimensional flows require accounting for the number and stability of limit cycles. Given that we do not record the presence of limit cycles here, in the main text we focus on systems with up to 2 fixed points of any type. For these systems, the precise number and type of limit cycle does not matter as much for the interpretation of the latent space. Systems with $(3,0)$ and $(0,3)$ topologies are shown in Fig.~\ref{fig:higher_fps}(b). $(3,0)$ systems are relatively rare and overlap with other (multi-)stable sytems. $(0,3)$ systems are classified either as excitable (as would be a system composed of a stable limit cycle with two unstable fixed points outside) or bistable (two stable limit cycles). 

\subsection{Latent space classifiers for topological structure prediction}
\label{sec:latent_classifiers}

\begin{figure*}
    \centering
    \includegraphics[scale=1]{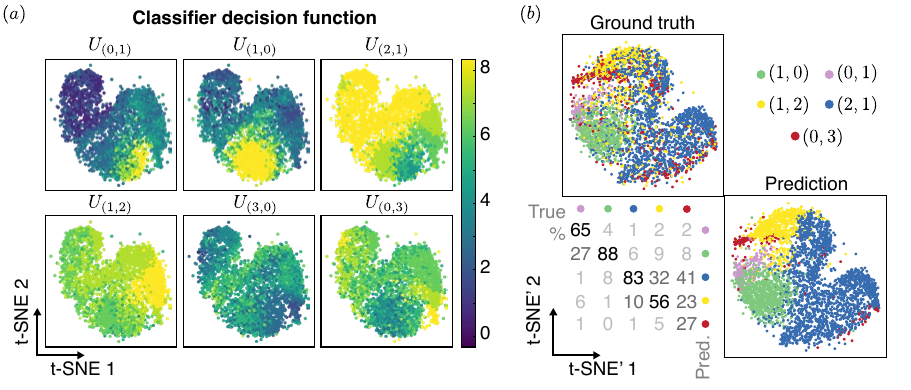}
    \caption{\textbf{Determination of latent space structure} \emph{(a)} Decision function of a multiclass linear SVM. Higher values mean higher probability of a point being in that class. \emph{(b)} Test performance of the classifier on additional $\sim 2500$ samples, restricted to the 4 classes of interest: misclassification mostly affects boundary points. Confusion matrix details percentages of misclassification: e.g. $27\%$ of oscillators are misclassified as monostable. Results restricted to the 5 most common categories with less than $3$ fixed points. For clarity, the $(3,0)$ points are omitted.}
    \label{fig:topology_latent_classifier}
\end{figure*}

To determine regions of latent space corresponding to different topological structures, we solve a supervised classification problem using a multiclass linear Support Vector Machine (SVM), which draws hyperplanes in the latent space to optimally discriminate between the different training data samples. We train our classifier on the pairs $(h_i, y_i)$ formed by the position in latent space $h_i$ of $5000$ examples and their ground truth topological structure $y_i=(n_\mathrm{stable}^i, n_\mathrm{unstable}^i)$ (Fig.~\ref{fig:topology_latent_classifier}(a-b)).

To obtain test statistics, we then compute the classification error (per-class and averaged) on an additional 5000 sample data point (Fig.~\ref{fig:topology_latent_classifier}(c)).

We repeat the process for different training and testing data with varying noise amplitude $D$.

\subsection{Visualization of latent space}
\label{sec:visualization}

The $128$-dimensional latent space encoded by $h$ is too large to be visualized directly. To understand variations in such high-dimensional spaces, we must use dimensional reduction techniques for visualization. Note that these projection techniques are solely for visualization: all clustering and distance metrics operate directly on the full 128-dimensional latent space.

Here we compare different projection methods and show that clustering in terms of fixed point topology is robust across visualization technique, indicating that it is not an artifact of the projection.

We compare different approaches here:
\begin{itemize}
    \item Principal Component Analysis (PCA) is a linear technique that projects data onto the latent-space directions where the data has the largest spread. It is interpretable and linear, but unfortunately is not expressive enough for data which has more than 2 or 3 natural axis of variations, or for which data clusters along curved manifolds.
    \item Multidimensional scaling (MDS) is a nonlinear method which attempts to map data points to a lower dimensional representation which preserves relative distances.
    \item t-distributed Stochastic Neighbors Embedding (t-SNE)  \cite{vandermaaten_tsne_2008} is a nonlinear technique that estimates a probability distribution that two samples are neighbors in high-dimensional space, and  distributes the samples in low-dimensional space such that the distribution in the low-dimensional space minimizes the Kullback-Leibler divergence between the two distributions with respect to the locations of the points in the map. It thus preserves local topological information (which samples neighbor each other), but can have difficulty with the global structure.
    \item Universal Manifold Approximation and Projection (UMAP) \cite{mcinnes_umap_2020} is a nonlinear technique that relies on the following assumptions: the data is uniformly distributed on a Riemannian manifold; the Riemannian metric is locally constant (or can be approximated as such); the manifold is locally connected. It tends to capture global information a little better than t-SNE, with the advantage that additional data points can be projected to the low-dimensional representation after fitting the learned manifold.
\end{itemize}

\subsection{Effect of transients on training and latent space}
\label{sec:transient_latent}

In this section, we investigate the effect of training and testing using data excluding initial starting transients. We expect classification to qualitatively change: in the absence of global flow information provided by the initial transients, the number of stable attractors visible in the input data and the transitions between them essentially determines the dynamical behaviors. 

\begin{figure*}
    \centering
    \includegraphics[width=.9\linewidth]{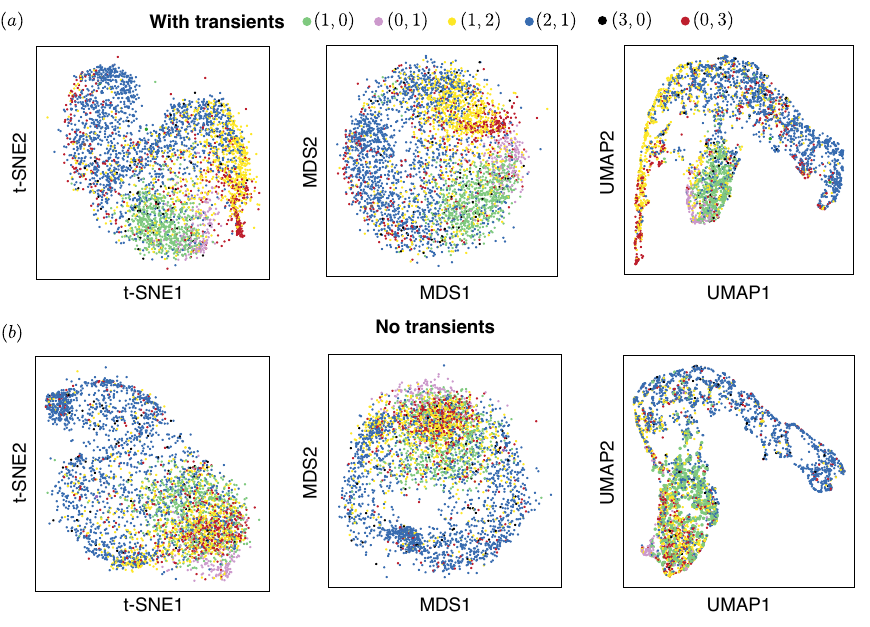}
    \caption{\textbf{Comparison of nonlinear dimensionality reduction for visualization and effect of transients on categorization.} \emph{(a)} For the network trained and run on data with initial transients ($\sim2300$ points) \emph{(b)} For the network trained and run on data without initial transients - excitable trajectories are harder to distinguish from monostable without wider exploration of the phase space.}
    \label{fig:comp_viz_transients}
\end{figure*}

Running these experiments, we find that mono- and bi-stability are still easily distinguished, but that excitable systems now reside `between' monostable and oscillatory systems (Fig.~\ref{fig:comp_viz_transients}(b)). This is in accordance with the traditional intuition that excitable systems are `nearly oscillators', or that oscillators are excitable system whose stable point has a vanishing stability basin~\cite{izhikevich_neural_2000}.

\section{Alternative initial conditions for training}
\label{sec:alternative_IC}

In this section, we explore the robustness of our results to alternative training data construction for the neural network. We use the same training ensemble as in Sec.~\ref{sec:trainingdata}, but instead of directly using trajectories starting at a randomly-chosen fixed point, we use a small sample $\{x_0^i(t), y_0^i(t)\}, i=1,\ldots,N_w$ of trajectories started at fixed points to estimate the boundaries of a region of interest for the initial conditions ($N_w = 10$). We then use the mean positions and standard deviations of these arrays $\langle x_0\rangle = 1/(N_tN_w)\sum_{i,t} x^i_0(t)$, $\sigma_{x,0}^2 = \langle x_0^2 \rangle - \langle x_0\rangle^2 $ to sample 30 initial conditions from the bivariate normal distribution $x\sim \mathcal{N}(\langle x_0\rangle,4\sigma_{x,0}^2)$, $y\sim \mathcal{N}(\langle y_0\rangle,4\sigma_{y,0}^2)$.

\begin{figure*}
    \centering
    \includegraphics[scale=1]{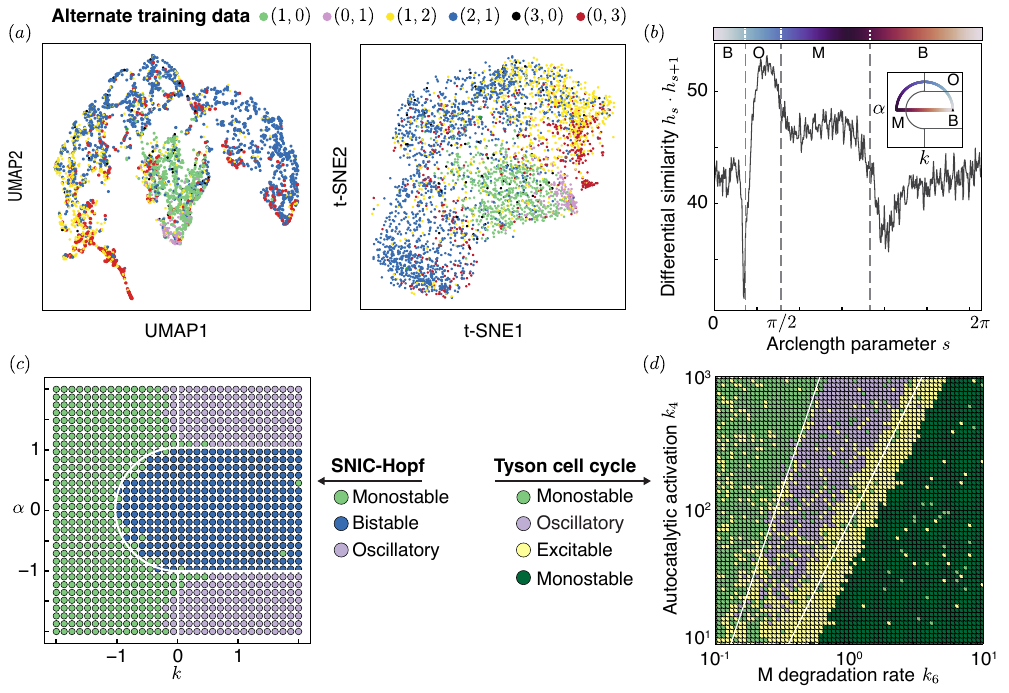}
    \caption{\textbf{Alternative initial conditions for training data.} \emph{(a)} UMAP and t-SNE projections of the latent space, colored by topological structure $(n_\mathrm{stab}, n_\mathrm{unstab})$. We recover the previous arrangements of the latent space. \emph{(b)} Variation in embedding for a traversal of SNIC-Hopf model space; we recover the behaviors at bifurcations. \emph{(c)} inferred SNIC-Hopf phase diagram from k-means clustering. \emph{(d)} Tyson cell cycle phase diagram from Agglomerative clustering. Previous results are recovered.}
    \label{fig:alternativeIC}
\end{figure*}

We then use the trained network on the previous test cases detailed above---results are shown in Fig~\ref{fig:alternativeIC}. We find that the newly-trained etwork performs comparably to the previous choice of initial conditions, showing the robustness of this training dataset.

\section{Comparison to an autocorrelation-based embedding}
\label{sec:autocorrelation}

To illustrate the advantages and limitations of our neural network-based approach, we compare our results to an approach using the auto-correlation matrix of trajectory segments \cite{takens_detecting_1981}.

We divide each trajectory $i=1,\ldots,N$ into segments $S^i_t = (\mathbf{x}^i_{t-n\tau}, \ldots,\mathbf{x}^i_{t-2\tau}, \mathbf{x}^i_{t-\tau}, \mathbf{x}^i_t)$ for $\tau =1$, $n=8$. From $N=30$ trajectories sampled at 100 time points, we thus obtain $3\times (100-8)$ samples of  $2n=16$-dimensional vectors. 
Defining the average over time points and trajectories $\langle \mathbf{x}^i_{t-j\tau}\rangle = N^{-1}N_t^{-1}\sum_{i,t} \mathbf{x}^i_{t-j}$, we compute the empirical covariance matrix
\begin{equation}
    C_{jk} = \left\langle( \mathbf{x}^i_{t-j\tau} - \langle \mathbf{x}^i_{t-j\tau}\rangle )(\mathbf{x}^i_{t-k\tau} - \langle \mathbf{x}^i_{t-k\tau}\rangle)\right\rangle_{i,t}
\end{equation}
and use the upper triangular part  $c = (C_{j\leq k})$ of this symmetric matrix to define a $2n(2n+1)/2=136$-dimensional vector as our data representation.

\begin{figure*}
    \centering
    \includegraphics[scale=1]{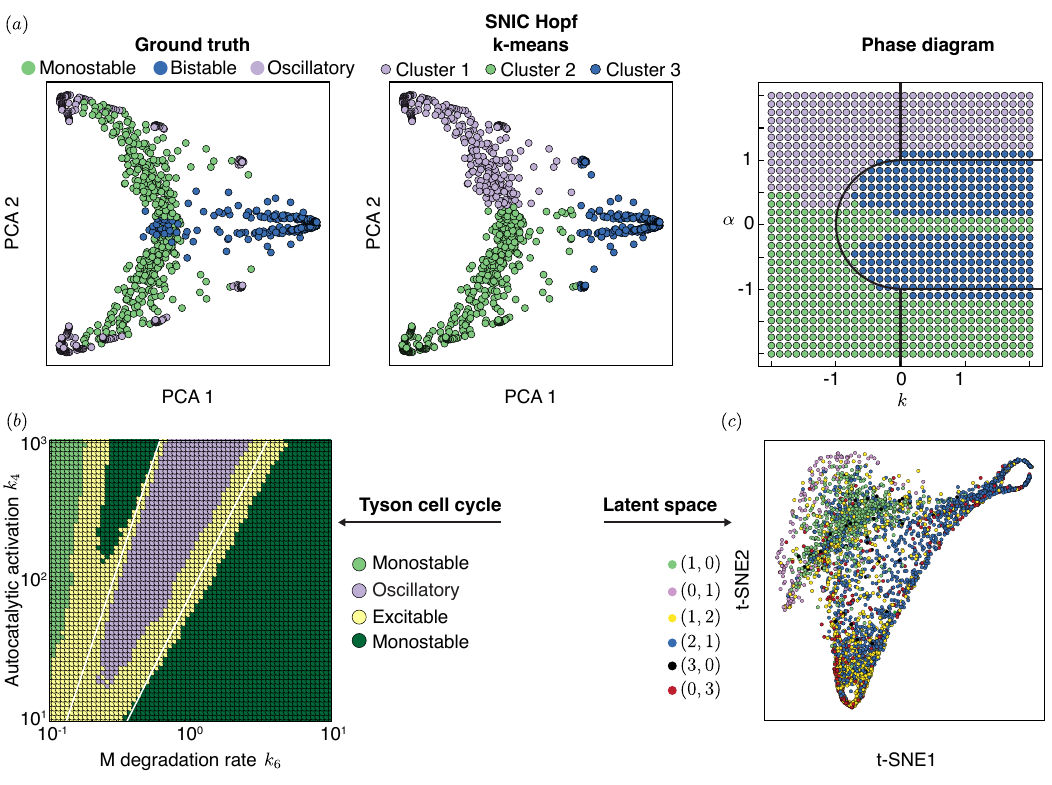}
    \caption{\textbf{Covariance of time delay-based characterization is simple, but less robust.} \emph{(a)} Test of the time-delay covariance method on the SNIC-Hopf test case: We find that while distinct dynamical behaviors are distinguishable (left), they are difficult to automatically cluster along behavioral regimes (center), leading to a less satisfactory phase diagram (right). \emph{(b)} Phase diagram of the Tyson cell cycle data using the time-delay method. Again, we partially recover the phase diagram. \emph{(c)} The latent space has some behavioral information, but topological regions are less separated.}
    \label{fig:PCAtimedelay}
\end{figure*}

We apply this method to the SNIC-Hopf, cell cycle and the neural network training data detailed above. This method can discriminate between some dynamical phases but is less robust (Fig.~\ref{fig:PCAtimedelay}(a-b)) --- strikingly, the PCA embedding of the correlation matrix reflects the symmetry of the system but distinct behaviors are scattered in sub-groups, making the clustering problem more difficult. The resulting latent space is also less structured than the one obtained from the neural network (Fig.~\ref{fig:PCAtimedelay}(c)). Although the time-delay method is simpler than our neural network-based approach, we thus see that this simplicity is at the expense of robustness and structure of the latent space.

\section{Nonlinearity detection in non-spiking excitable systems}\label{sec:nonlinear_detection}

To study the possibility of detecting nonlinearities in the absence of visible strongly-nonlinear behaviors, in the main text Fig.~4 we consider the system 
\begin{subequations}
    \begin{align}
        \mathrm{f}_x & = -x + \nu x^2 - x^3 - y, \\ 
        \mathrm{f}_y & =  \epsilon(x - y).
    \end{align} %
\end{subequations}
This system can exhibit excitable behavior if $\nu > 2$: If $\epsilon \rightarrow 0$, starting at $(0,0)$ the slow $y$-dynamics can be ignored and this system reduces to the one-dimensional cubic system
\begin{align}
\dot{x} =  - x + \nu x^2 - x^3
\end{align} 
This double well-type system has three fixed points: one stable at the origin, one unstable, and another stable one further along the $x$-axis.  The position of the unstable fixed point defines the threshold $\delta$ for excitability, while the right-most stable one sets the maximum range $d_\mathrm{max}$ of the excursion before the slow $y$-dynamics bring the system back to the origin.

We then solve the SDEs
\begin{equation}
    \mathrm{d}\mathbf{x} = \mathbf{f}(\mathbf{x})\mathrm{d}t + \Sigma\cdot\mathrm{d}\boldsymbol{\zeta}
\end{equation}
with diagonal anisotropic noise
\begin{equation}
    \Sigma = \begin{pmatrix}
        \sqrt{2D} & 0 \\ 0 & \sqrt{2D\epsilon}.
    \end{pmatrix}
\end{equation}
The linear part of the flow is 
\begin{align}
    \mathbf{f}_{lin}(\mathbf{x}) = \begin{pmatrix}
        -1 & - 1 \\ \epsilon & -\epsilon 
    \end{pmatrix} \begin{pmatrix}
        x \\ y
    \end{pmatrix}
\end{align}
The stationary dynamics of the linear system are exactly solvable: with standard Gaussian white noise, the resulting trajectories are Gaussian-distributed with mean $0$, and components are fully determined by the steady-state autocorrelation functions
\begin{subequations}
\begin{align}
    \langle \mathbf{x}(\omega)\otimes \mathbf{x}(\omega')\rangle & = \begin{pmatrix}
        \langle x(\omega) x(\omega')\rangle & \langle x(\omega) y(\omega')\rangle \\ 
        \langle y(\omega) x(\omega')\rangle & \langle y(\omega) y(\omega')\rangle
    \end{pmatrix} \\ 
    & = \frac{2D}{(2\epsilon-\omega^2)^2+(1+\epsilon)^2\omega^2}\delta(\omega+\omega') \begin{pmatrix}
        \omega^2+\epsilon(1+\epsilon) & (1+\epsilon)i\omega+\epsilon(\epsilon-1)\\
        -(1+\epsilon)i\omega +\epsilon(\epsilon-1)  & \epsilon\omega^2 +\epsilon(1+\epsilon)
    \end{pmatrix}.
\end{align}
\end{subequations} %
Given the Gaussian nature of linear trajectories $\mathbf{x}_\mathrm{lin}(t)$, statistical tests of multivariate normality are able to detect the presence of non-linearity which generically leads to non-vanishing higher-order moments.

\section{Regions of latent space inaccessible to gradient dynamics}\label{sec:gradient}

In this section we study the differences in latent space occupation between systems drawn from the random ensemble of cubic polynomial systems.

\begin{figure*}
    \centering
    \includegraphics[scale=1]{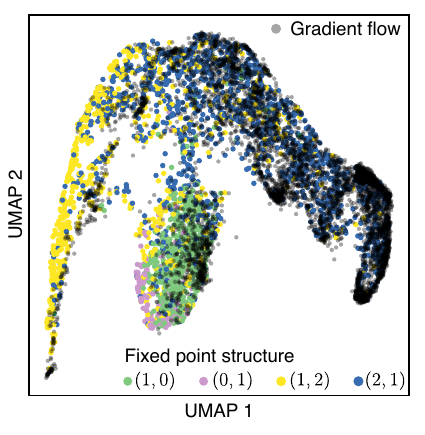}
    \caption{\textbf{Behaviors inaccessible to gradient flow.} Gradient flow systems, in gray, are constrained in what behaviors they can display. Their positions in latent space reflect these constraints: they cannot be oscillatory nor excitable, and even certain monostable behaviors, such as non-normal linear response, are not accessible.}
    \label{fig:conservative_vs_ensemble}
\end{figure*}

Consider a system obeying gradient flow, meaning that its deterministic dynamics are given by the gradient of a potential function
\begin{equation}
    \frac{\mathrm{d}\mathbf{x}}{\mathrm{d}t} = - \nabla V(\mathbf{x})
\end{equation}
In the language of dynamical systems, this system derives from a Lyapunov functional. Our random ensemble of systems is a superset of such gradient systems: indeed, rewriting our dynamics using a tensor formalism $\mathbf{x}=x_i \mathbf{e}_i$ and using the Einstein summation convention, the differential equations read
\begin{equation}
    \frac{\mathrm{d}x_i}{\mathrm{d}t} = A_{ij}x_j + B_{ijk}x_j x_k + C_{ijkl} x_j x_k x_l.
\end{equation}
Any dynamics whose linear part $A$ for instance is not symmetric cannot derive from a potential, and similarly if higher-order terms break circular permutation symmetry such as if $B_{ijk} \neq B_{kij} \neq B_{jki}$,  $C_{ijkl} \neq C_{lijk}$, and so on.

To construct an ensemble of dynamical systems restricted to planar cubic dissipative systems, we consider potential functions expanded onto fourth-order polynomials
\begin{equation}
    V(x,y) = v_{20}x^2 + v_{11}xy  + v_{02}y^2 + v_{30}x^3 + v_{21}x^2 y + v_{12} x y^2 + v_{03} y^3 + v_{40} x^4 + v_{31} x^3 y + v_{22}x^2 y^2 + v_{13} x y^3 + v_{04} y^4.  
\end{equation}
Note that constant and linear terms are omitted, since they can be eliminated by change of energy scale and coordinate origin respectively. To impose boundedness conditions similar to the ones derived in Eqs.~\eqref{eq:stab_conditions_ensemble}, the conditions $a_{30} <0, b_{03} <0$ impose
\begin{subequations}
    \begin{align}
        v_{40} & > 0 \\
        v_{04} & > 0,
    \end{align}
\end{subequations}
while the conditions $a_{21} = -b_{30}$ and $a_{03} = -b_{12}$ impose $v_{31} = 0$ and $v_{13}=0$ respectively. The condition $a_{12}=-b_{12}$ finally imposes $v_{22} = 0$. As before, in practice we randomly sample all $v_{ij}$ from the unit normal distribution, replace $v_{40} \leftarrow |v_{40}|, v_{04} \leftarrow |v_{04}|$, and finally set $v_{31}=v_{13} = v_{22} = 0$.

Simulating $5000$ systems, we find that as expected potential systems can reproduce many, but not all, behaviors seen in the original ensemble. In particular, potential systems cannot produce neither oscillators nor excitable systems (Fig.~\ref{fig:conservative_vs_ensemble}).

\section{Characterization by comparison to reference systems}
\label{sec:refsystems}

\begin{figure*}
    \centering
    \includegraphics[scale=1]{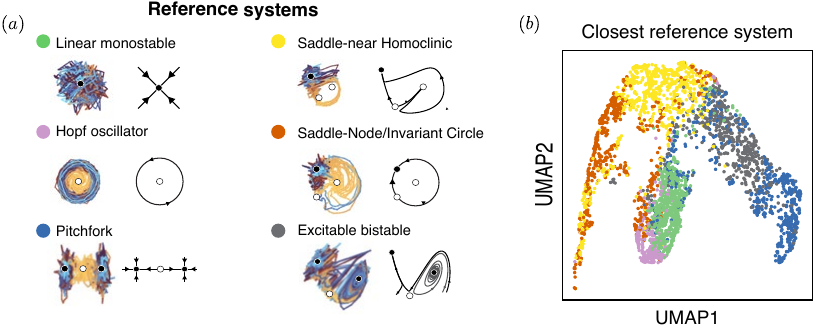}
    \caption{\textbf{Finer dynamical characterizations though comparison.} \emph{(a)}  A set of hand-picked canonical systems provides reference points to compare dynamics against. \emph{(b)} Coloring the latent space by closest reference system provides a finer characterization than fixed point topology.}
    \label{fig:beyond_topology}
\end{figure*}

We saw that the characterization of the global geometric features of the deterministic flow of dynamical systems by the number and type of its fixed points can miss important geometric features, such as the possibility of spiking trajectories, as can happen in the FitzHugh-Nagumo model in the excitable regime (Sec.~\ref{sec:fhn}).
To provide a characterization accounting for geometrical features, inspired by the notion of canonical systems \cite{izhikevich_neural_2000}, we compare trajectory data to data generated from a selection of reference systems to draw finer regions in latent space. 
We thus introduce a range of reference behaviors, namely systems described by standard normal forms and examples of geometrically distinct variations of excitable and bistable systems (Fig.~\ref{fig:beyond_topology}(a)). Coloring points in latent space by a measure of distance in latent space, we find distinct regions corresponding to these subtypes of systems - a more refined version of the dynamical `phase diagram' established by topology.

Below, we define the six systems we will use as reference, then detail the construction of the associated regions in latent space.

\subsection{Reference dynamical systems}

The reference systems are common normal forms or realizations of canonical models from \cite{izhikevich_neural_2000}. Given that we focus on bounded flows in the plane, our systems have topological index $1$ \cite{strogatz_nonlinear_2024}. To cover possible systems which have the same topological index (limiting ourselves to systems with at most one limit cycle, and no limit cycles with another stable fixed point), we consider 6 different dynamical systems:
\begin{enumerate}
    \item[1-] a system with only one stable fixed point,
    \item[2-] one oscillatory system with one unstable fixed point, 
    \item[3,4-] two bistable systems whose stable points have different geometries of their basins of attraction,
    \item[5,6-] two excitable systems realizing different canonical models of \cite{izhikevich_neural_2000}.
\end{enumerate}

\subsubsection{Monostable system}

As our reference stable system with one single stable hyperbolic fixed point we use a simple stable linear system
\begin{subequations}
    \begin{align}
        \dot{x} & = -x, \\
        \dot{y} & = -y. 
    \end{align} \label{eq:ref_mono}
\end{subequations}
For isotropic noise of variance $D$, the steady-state probability of finding this potential system at a radius $r$ away from the origin is Gaussian with variance $D$, namely
\begin{equation}
    p(r) \sim e^{-\frac{r^2}{2D}}.
\end{equation}

\subsubsection{Hopf oscillator}

Our reference limit cycle oscillator will be the one stemming from the normal form of the supercritical Hopf bifurcation in Cartesian coordinates
\begin{subequations}
    \begin{align}
        \dot{x} & = x - y - (x^2+y^2)x \\
        \dot{y} & = x + y - (x^2+y^2)y
    \end{align} \label{eq:ref_hopf}
\end{subequations}
In polar coordinates, the dynamics read as 
\begin{subequations}
    \begin{align}
        \dot{r} & = r - r^3, \\
        \dot{\theta} & = 1.
    \end{align} \label{eq:ref_hopf_polar}
\end{subequations}
which make visible that the circle of unit radius around the origin is a limit cycle with angular frequency $1$ in the absence of noise.

\subsubsection{Pitchfork bistable system}

We consider the normal form of a supercritical pitchfork to obtain a reference bistable system. This is the same systems as the one used in section~\ref{sec:noise_effects} with $r=1$.
\begin{subequations}
    \begin{align}
        \dot{x} & = x - x^3 \\
        \dot{y} & = -y
    \end{align} \label{eq:ref_bistable}
\end{subequations}
The two stable fixed points are located at $(x=\pm1,y=0)$.

\subsubsection{Saddle-Node on Invariant Circle}

As an example of a class-$1$ excitable system with 1 stable, 2 unstable fixed points, we consider the Saddle-Node on Invariant Circle (SNIC) system in its excitable regime \cite{izhikevich_neural_2000}.
In polar coordinates $(r,\theta)$, the model reads
\begin{subequations}
    \begin{align}
        \dot{r} & = r-r^3 \\
        \dot{\theta} & = \Omega +r\sin\theta
    \end{align} \label{eq:ref_snic_polar}
\end{subequations}
with $|\Omega|<1$ in the excitable regime (when $|\Omega|>1$, the system is an oscillator). In all results shown here, we use $\Omega=0.8$.
In Cartesian coordinates $x=r\cos\theta, y=r\sin\theta$, the dynamics are given by
\begin{subequations}
\begin{align}
    \dot{x} = & x- (x^2+y^2)x -\Omega y - xy \\
    \dot{y} = & y- (x^2+y^2)y +\Omega x + x^2
\end{align} \label{eq:ref_snic}
\end{subequations}
The system has fixed points at the origin and at $(-\Omega, \pm\sqrt{1-\Omega^2})$. The stable fixed point is at $(-\Omega, \sqrt{1-\Omega^2})$.

\subsubsection{Saddle Homoclinic Orbit excitable system}

As another canonical model of excitability, with 1 stable, 2 unstable fixed points, but this time of class-$2$ excitability \cite{izhikevich_neural_2000}, we consider the system presented in section~\ref{sec:SHOsystem} in the monostable excitable regime
\begin{subequations}
\begin{align}
    \dot{x} = & -x + \nu x^2 -x^3 - Ay \\
    \dot{y} = & \epsilon\left(\frac{x^2}{2} -\frac{x}{2} -y\right)
\end{align} \label{eq:ref_sho}
\end{subequations}
with $A=7$, $\nu=4$, and $\epsilon=1$. This system can be considered a variant of the Saddle-Homoclinic orbit system introduced in \cite{izhikevich_neural_2000}. Its nullclines are given by
\begin{subequations}
\begin{align}
    y = & (-x + \nu x^2 -x^3)/A \\
    y = & \frac{x^2}{2} -\frac{x}{2}
\end{align}
\end{subequations}
which intersect at $(0,0)$ and for $x_\pm=\frac{\nu}{2}-\frac{A}{4} \pm \frac{1}{2}\sqrt{\left(\frac{A}{2}-\nu\right)^2 + 2A - 4}$. For $\epsilon=1$, the fixed point at $x=x_+$ is unstable. 

\subsubsection{Excitable bistable system}
We use the same system as Eq.~\eqref{eq:ref_sho}
\begin{subequations}
\begin{align}
    \dot{x} = & -x + \nu x^2 -x^3 - Ay \\
    \dot{y} = & \epsilon\left(\frac{x^2}{2} -\frac{x}{2} -y\right)
\end{align} \label{eq:ref_sho_bis}
\end{subequations}
but with a different timescale $\epsilon=4$, all other parameters being the same ($A=7, \nu=4$). With $\epsilon=4$, this system is bistable, with both fixed points at $x=x_\pm$ stable. 

\subsection{Barycentric coordinate system}

To establish a comparison between simulated random dynamical systems and our reference systems, we introduce planar barycentric coordinates which allow us to define a ``closest'' reference system to any given point in the latent space.

\begin{figure*}
    \centering
    \includegraphics[width=0.5\linewidth]{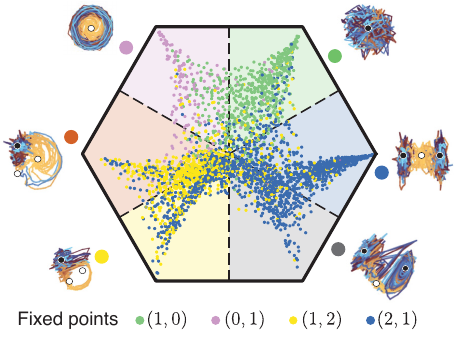}
    \caption{\textbf{Latent space in barycentric coordinates.} Using reference systems to construct a representation of the systems allows to distinguish sub-behaviors: for instance, bistable systems split themselves between pitchfork, excitable bistable, and SHO systems. Counterclockwise, from rightmost vertex: Pitchfork, monostable, Hopf, SNIC, SHO, Excitable-Bistable.}
    \label{fig:barycentric}
\end{figure*}
For $\alpha=0,\ldots,5$, we place vertices at position
\begin{equation}
    \mathbf{b}_\alpha = \begin{pmatrix}
        \cos \frac{2\pi\alpha}{6} \\
        \sin \frac{2\pi\alpha}{6}
    \end{pmatrix}
\end{equation}
with each vertex corresponding to a reference system. We then place systems on the plane spanned by the $\mathbf{b}_\alpha$ according to their Euclidean distance $d_\alpha$ in latent space to each reference system at position 
\begin{align}
    \mathbf{r} = \frac{\sum_\alpha d_\alpha^{-4} \mathbf{b}_\alpha}{\sum_\beta d_\beta^{-4}}
\end{align}
such that all points are inside the convex hull defined by the vertices. A system very similar to a given reference system $d_\alpha \approx 0$ would be mapped to the vertex $\mathbf{b}_\alpha$ itself. We then assign systems to the closest reference system in this barycentric coordinate plane (Fig.~\ref{fig:barycentric}).

\section{Contrastive learning for run-and-tumble bacterial track characterization}

In this section, we detail the application of the contrastive approach to \emph{E.\ coli} tracks displaying run-and-tumble motion. We use the data from Dufour et al.\ \cite{dufour_direct_2016}. The data consists of a control set of approximately 13000 wild-type \emph{E.\ coli} tracks, and a smaller set of $\sim$$4000$ trajectories of bacteria genetically engineered to fluorescently report the presence of the proteins $\text{CheB}$ and $\text{CheR}$ known to be involved in motility regulation. These mutants are also engineered to have a wider range of $\text{CheB}$ and $\text{CheR}$ concentrations than wild type \emph{E.\ coli}.

After obtaining finite-difference estimates of the velocity $\left(v_x(t),v_y(t)\right)$ from the trajectories $\left(x(t), y(t)\right)$, we train a neural network on the velocity time series of the control data with augmentations in the orthogonal group $O(2)$. Unlike the neural networks used in other sections, we only have a single trajectory per system and thus do not need the permutation of trajectory invariance. We then resample trajectories by linear interpolation to a common number of time points ($400$) to maintain a constant input length across data samples. Those interpolated trajectories are centered and scaled by their standard deviation (standardized) before passing them into the neural network. Using a three-layer MLP with the same contrastive loss and optimization scheme as SI Sec.~\ref{sec:contrastive_learning} and a 64-dimensional latent space and batch size of 2000, after $2\cdot10^4$ epochs the loss converged to its final value. We note that this normalization means that the neural network output is blind to the overall velocity scale. We also remark that trajectories can cover different time spans, leading this neural network to also be blind to overall time scale: while this scale-invariance prevents the distinction of systems which only differ by overall speed, it is still relevant to the observables of \cite{dufour_direct_2016}, in which the analysis is mainly based on ratios of time scales. 
%
%
We note that despite the velocity scale invariance, the effective translational diffusion coefficient still inversely correlates with tumble bias as expected.

\begin{figure*}
    \centering
    \includegraphics[scale=1]{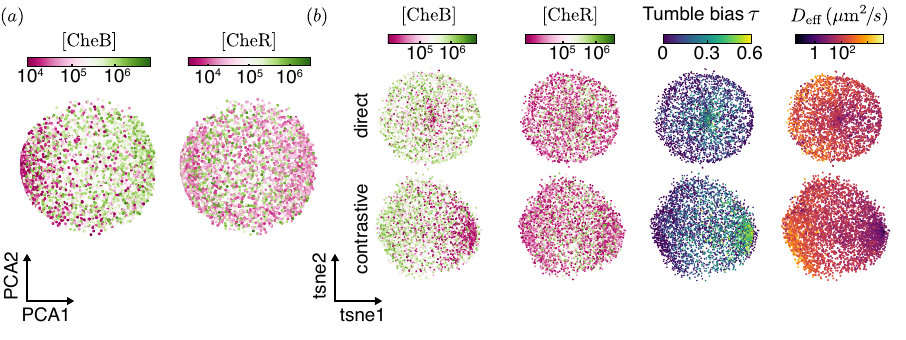}
    \caption{\textbf{Contrastive cartography for symmetry-informed dimensional reduction.} \emph{(a)} PCA projection of $\text{[CheB]}$, $\text{[CheR]}$, showing the asymmetric correlations of  the proteins with tumbling bias. \emph{(b)} Comparison of direct t-sne projection of velocity tracks and projection of the contrastive latent space. By building in invariance to orthogonal transformations, the contrastive approach provides better linear separability.}
    \label{fig:SIrunNtumble}
\end{figure*}

\bibliography{biblio}